\documentclass{article}
\usepackage[utf8]{inputenc}
\usepackage{cite}
\usepackage{authblk}
\usepackage{hyperref}
\usepackage{amsmath}
\usepackage{leitian_v2}
\usepackage{amssymb}
\usepackage[margin=1.3in]{geometry}
\usepackage{graphicx}
\usepackage{upgreek}

\usepackage{amsfonts}
\usepackage{textcomp}
\usepackage{comment}
\usepackage{array}
\usepackage{multirow}
\usepackage[table]{xcolor}
\usepackage{caption} 
\usepackage{booktabs}
\usepackage[ruled, lined, commentsnumbered, longend]{algorithm2e}

\newcommand{\tn}[1]{\textnormal{#1}}
\newcommand{\tb}[1]{\textbf{#1}}
\newcommand{\mb}[1]{\mathbf{#1}}
\newcommand{\mc}[1]{\mathcal{#1}}

\title{Adaptive 3D descattering with a dynamic synthesis  network}

\author[1,$\star$]{Waleed Tahir}
\author[1,$\star$]{Hao Wang}
\author[1,2,*]{Lei Tian}

\affil[1]{Department of Electrical and Computer Engineering, Boston University, Boston, MA 02215, USA.}
\affil[2]{Department of Biomedical Engineering, Boston University, Boston, MA 02215, USA.}
\affil[$\star$]{Equal contribution.}
\affil[*]{Correspondence: leitian@bu.edu, Tel.: 1-617-353-1334}

\begin{document}

\maketitle

\clearpage
\begin{abstract}
Deep learning has been broadly applied to imaging in scattering applications.
A common framework is to train a descattering network for image recovery by removing scattering artifacts. 
To achieve the best results on a broad spectrum of scattering conditions, individual ``expert'' networks need to be trained for each condition. 
However, the expert's performance sharply degrades when the testing condition differs from the training.
An alternative brute-force approach is to train a ``generalist'' network using data from  diverse scattering conditions. 
It generally requires a larger network to encapsulate the diversity in the data and a sufficiently large training set to avoid overfitting.
Here, we propose an {\it adaptive learning} framework, termed dynamic synthesis network (DSN), which {\it dynamically} adjusts the model weights and {\it adapts} to different scattering conditions. 
The adaptability is achieved by a novel ``mixture of experts'' architecture that enables dynamically synthesizing a network by blending multiple experts using a gating network.
We demonstrate the DSN in holographic 3D particle imaging for a variety of scattering conditions.
We show in simulation that our DSN provides generalization across a {\it continuum} of scattering conditions.
In addition, we show that by training the DSN entirely on simulated data, the network can generalize to experiments and achieve robust 3D descattering.
We expect the same concept can find many other applications, such as denoising and imaging in scattering media.
Broadly, our dynamic synthesis framework opens up a new paradigm for designing highly {\it adaptive} deep learning and computational imaging techniques.
\end{abstract}

\section{Introduction}

Deep learning (DL) has become a powerful technique for tackling complex yet important computational imaging problems~\cite{barbastathis2019use}, such as phase imaging~\cite{sinha2017,xue2019,wang2020,matlock2021p}, tomography~\cite{wang2020deep,liu2020rare,wu2020simba,gupta2018cnn}, ghost imaging~\cite{wang2019learning,rizvi2020ghost,lyu2017deep,li2020compressive}, lightfield microscopy~\cite{wagner2021deep, wang2021real}, super-resolution imaging~\cite{wang2019deep, liu2019deep, rivenson2017deep}, enhancing digital holography~\cite{rivenson2018phase,ren2019end, rivenson2019deep,wu2019bright} and imaging through scattering media~\cite{li2018deep,sun2019image,li2021, li2018imaging}.
Within these computational imaging applications, one of the prevalent problems is ``descattering'', or removing scattering artifacts.
For this purpose, a deep neural network (DNN) is generally trained to perform descattering, either directly on the measurement~\cite{wang2020,xue2019,sinha2017,wang2019learning,ren2019end,rivenson2018phase, li2021,li2020compressive,li2018deep} or on the object-space projection~\cite{rizvi2020ghost,lyu2017deep}. 
Alternative to this ``end-to-end'' framework, another approach is to employ a pretrained DNN as the learned prior in an iterative model-based reconstruction algorithm to progressively mitigate scattering artifacts~\cite{liu2020rare,wu2020simba}.
\\

While increasingly effective, the existing descattering DL frameworks are fundamentally impeded by an outstanding challenge. 
They generally demonstrate optimal performance only when the scattering condition in the testing data match well with the training data, and the performance  degrades sharply when the scattering conditions are mismatched~\cite{li2018deep,li2021}.
Thus, if a task requires working with many different cases of scattering, it generally needs to train multiple ``expert'' networks, each optimized for a specific scattering condition~\cite{li2020compressive,sun2018ef,gupta2018cnn}.
Such expert networks are very limiting since they require {\it a priori} knowledge of the scattering condition for optimally training the DNN.
An alternative ``brute-force'' method is to train a single ``generalist'' network using a larger data set combining diverse scattering conditions~\cite{li2018deep,li2021}.
However, this generalist approach often achieves lower performance than the expert since the generalist needs to perform the challenging task of extracting generalizable features across different scattering cases.
\\

Given these limitations, it is highly desirable to develop a unified {\it adaptive} DL framework that can robustly handle a broad range of scattering conditions.
While this challenge has not been addressed broadly in the literature, one approach~\cite{sun2019image} is to train a bank of expert DNNs, each for a different scattering condition. This bank is preceded by a separately trained classification DNN, whose purpose is to select a suitable descattering expert based on the input data at the test time.
However, this architecture suffers from poor scalability, and its overall performance is fundamentally limited by the disjoint classification and descattering networks, which are separately trained and applied to the input data during the testing.
\\

Another inspiration stems from the framework of ``mixture of experts (MoE)'', in which a mixture of expert networks is fused into a single network to achieve better generalization~\cite{6215056,agostinelli2013adaptive, 8663454, yang2020multi}. 
In image denoising, a MoE framework exploits a weighted sum of the output from different expert denoising DNNs.
The weights for combining the experts are obtained either by solving a separate nonlinear optimization problem or training a separate DNN. 
This approach has shown significant improvement to the denoising DNN's robustness against different noise types (e.g. Poisson vs Gaussian) and noise levels~\cite{8663454, agostinelli2013adaptive}.
In phase retrieval, a ``learning to synthesize'' MoE framework is employed to fuse low- and high- spatial frequency information extracted separately from two expert networks and later synthesized by a residual learning approach~\cite{deng2020learning}. 
This framework enables phase recovery with improved spatial resolution and resilience to high-noise conditions.
\\

\begin{figure}
	\centering
	\includegraphics[width=1\linewidth]{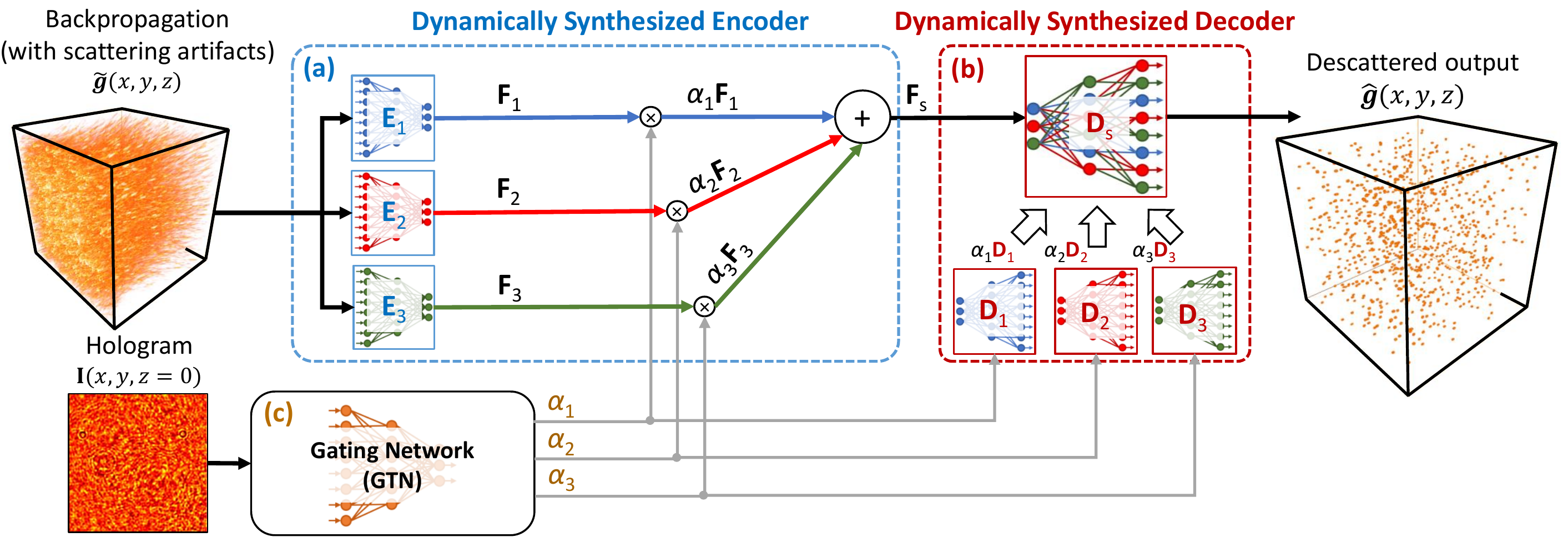}
	\caption{\textbf{Dynamic synthesis network (DSN) framework.} 
		The DSN combines multiple DNNs for adaptively removing scattering artifacts in the input. 
		(a) In the first stage, the expert encoders $\tn{E}_{i}$ ($i\in\{1,2,3\}$) extract a diverse set of multi-scale spatial features from the holographically back-propagated input volume $\tilde{\mb{g}}$. 
		The extracted multi-scale feature maps from each encoder are labeled as $\tb{F}_{i}$. 
		To adaptively process an input with an arbitrary scattering condition, a dynamically synthesized feature map $\tb{F}_{s}$ is computed as a weighted sum of the expert feature maps: $\tb{F}_{s} = \sum_{i=1}^3 \alpha_i\tb{F}_{i}$.
		The synthesized feature $\tb{F}_{s}$ is fed into a dynamically synthesized decoder $\tn{D}_{s}$ to produce the descattered output $\hat{\mb{g}}$.
		(b) Different from the encoder, the decoder $\tn{D}_{s}$ is computed as a weighted sum of the expert decoders' network parameters: $\tn{D}_{s} = \sum_{i=1}^3 \alpha_i\tn{D}_{i}$.
		(c) The GTN provides the adapting mechanism by predicting the synthesis weights $\alpha_{i}$ based on the matching hologram input.
		}
	\label{fig:overview}
\end{figure}

In this paper, we present the first, to the best of our knowledge, holistic {\it adaptive} descattering DL framework.
The proposed DNN, termed dynamic synthesis network (DSN), further advances the MoE framework by a novel expert mixing scheme.  
At the high-level, the DSN consists of multiple expert descattering networks for learning several sets of object/scattering features to achieve optimal descattering.
A gating network (GTN) is introduced to adaptively make ``central cognitive decisions'' for mixing the experts into a unified synthesized network (Fig.~\ref{fig:overview}). 
Once trained, the DSN adapts to each input during the inference and computes the optimal weights and the resulting synthesized network ``on-the-fly'', hence achieving   
``end-to-end'' adaptive descattering.
Instead of directly mixing the experts' output in the existing MoE models, the DSN architecture adaptively adjusts the network's (internal) parameters.
Specifically, the DSN works by {\it continuously} mixing the features maps extracted by different experts to synthesize an optimal feature representation of the input in a high dimensional feature space (Fig.~\ref{fig:overview}). 
We show that this novel structure allows DSN to perform adaptive descattering in a {\it continuum} of scattering levels and achieve superior performance across a broad range of scattering conditions in simulation and experiments.
%
%
\\

In this work, we use 3D inline holography as the testing bed to demonstrate our DSN framework.
Specifically, we perform 3D descattering of volumetric particle field reconstructions from the single-shot inline holography.
We demonstrate the adaptability and robustness of the DSN over a broad range of scatterer densities, sizes and refractive index contrast in simulation and further validate the DSN on experimental measurements with different scatterer densities. 
We show that the DSN can adaptively remove scattering artifacts even if the scattering condition has never been ``seen'' during the training, and its performance is comparable to, if not even better than, the expert network separately trained at the matching scattering condition.
We show that the DSN provides better generalization performance than the generalist network that is trained on multiple scattering densities simultaneously.
\\

A widely perceived challenge in demonstrating the effectiveness of a supervised DNN model in real optical experiments is the difficulty in capturing a sufficiently large data set with paired input and ground-truth data to train or fine-tune the DNN.
This is indeed the case in our intended application, which involves imaging freely moving micron-sized particles in fluid~\cite{katz2010applications}.
To overcome this challenge, we apply our recently developed multiple-scattering simulator trained DNN framework~\cite{matlock2021p}. 
Specifically, we train the DSN {\it entirely using simulated data}, which are generated by the beam-propagation model~\cite{wang2021}, in place of physically acquiring paired training data.
We show that this {\it simulator-trained} DSN presents similar behaviors on the experimental data to those observed in simulation and can robustly perform adaptive 3D descattering across a broad range of scatterer densities.
By doing so, we verify that the DSN can be successfully translated from simulation to real experiment.
We also highlight the synergies between the DSN framework and the physics-based multiple-scattering simulation to enable scientific and biomedical applications where the ground truth is hard to obtain.
\\

Overall, our contribution is a novel adaptive DL framework that achieves generalizable performance across a broad range of scattering conditions using a single holistic DNN architecture.
We demonstrated this framework on 3D particle imaging using inline holography with different particle densities, particle sizes and refractive index contrast.
We expect that the same dynamic synthesis framework can be adapted to many other imaging applications, such as image denoising~\cite{weigert_content-aware_2018}, imaging in dynamic scattering media~\cite{wang2021non,zheng2021incoherent}, computational fluorescence microscopy~\cite{pegard2016compressive,xue2020single}, and imaging~\cite{li2018deep,sun2019image,li2020compressive,li2021} and light control~\cite{turpin2018light, rahmani2020actor, turpin2020spatial, skarsoulis2021p} in complex media, such as biological tissues.
Broadly, our dynamic synthesis framework opens up a new paradigm for designing highly {\it adaptive} DL-based computational imaging techniques.

\section{Results}
In this section, we provide an overview of the DSN framework, followed by results on simulation and experiments.

\subsection{The DSN Framework}
\label{DSN_explaination}
The unique properties of the DSN include its dynamically synthesized feature representations of the input and the adaptively tuned network parameters, both of which are adjusted ``on-the-fly'' at each inference time to achieve adaptation.
This is in stark contrast to conventional DNNs, which  perform direct inference with pre-trained ({\it i.e. fixed}) network parameters. 
The DSN also bypasses the limitations in rigid model switching between a {\it fixed} set of expert DNNs, which makes the DSN more versatile and scalable.
The DSN enables synthesizing a network in a {\it continuous} high-dimensional feature space so that it can provide optimal performance in a {\it continuum} of scattering conditions.
The adaptability of the DSN stems from the interplay between a gating network (GTN) and a consortium of expert DNNs.
Each expert DNN in the DSN extracts certain spatial features to provide a diverse representation of the input.
The GTN provides the feedback needed to intelligently and dynamically fuse the extracted features and synthesize the feature representation tailored to the current input, as further illustrated in Fig.~\ref{fig:feature}.
It also enables computing the optimal mixture of the network parameters to utilize the synthesized features for descattering at different conditions.
%
%
\\

In this proof-of-concept study, we demonstrate the utility of our DSN framework to perform 3D descattering on holographically back-propagated 3D volumes containing high-contrast densely distributed particles (see details in Section~\ref{sec:setup}).  
Due to the large scale of the problem, we choose three experts to remove scattering artifacts from the backpropagated volumes, in order to strike a balance between the descattering performance and the computational cost.
The effects of the number of expert networks are further studied in Fig.~\ref{fig:s_2v3}.
\\

The schematics of our DSN framework is shown in Fig.~\ref{fig:overview}.
The input to the network is a preprocessed scattering-contaminated backpropagated volume from the hologram (see details in Section~\ref{sec:holo-prop} and Section~\ref{sec:preprocess}).
The network is trained to remove scattering artifacts, whose severity is highly dependent on the scatterer density, size, refractive index contrast, and further complicated by the depth-varying characteristics throughout the 3D volume~\cite{tahir2019,wang2021}.
Within the DSN, each expert has the same modified V-net structure~\cite{milletari2016}, which is further split into the expert encoder and expert decoder for performing dynamic synthesis (see Fig.~\ref{fig:overview}, and more details about the network structure in Section~\ref{sec:network}, Fig.~\ref{fig:s_Vnet} and Fig.~\ref{fig:s_exp_struc}).
\\

In the first stage, each expert encoder independently extracts a set of multi-scale spatial feature maps from the scattering-contaminated input volume.
Since each expert encoder has different ``specializations'', the combined set of feature maps provide a diverse representation of the scattering volume that provides the basis for adaptation.
To intelligently utilize these multi-scale features for processing an arbitrary scattering condition, a linearly weighted sum of the extracted feature maps is computed using the ``synthesis weights'' predicted by the GTN at each inference time (see Fig.~\ref{fig:overview}(a)). 
By doing so, this dynamically synthesized encoder provides a set of tailored feature maps to represent the input (see an illustration in Fig.~\ref{fig:feature}(a)). 
The training of the DSN effectively optimizes the set of bases (feature maps) in order to get a more generalizable representation for different scattering conditions. 
\\

On the decoding end, the DSN dynamically synthesizes the decoder during the inference by mixing a set of expert decoders.
This process is performed by directly computing a linearly weighted sum of the network parameters of the expert decoders  (see Fig.~\ref{fig:overview}(b)).
Finally, the synthesized features from the encoder are processed by the synthesized decoder to produce the descattered output volume (see an illustration in Fig.~\ref{fig:feature}(b)).
\\

The GTN provides the ``feedback signal'' for dynamic synthesis by predicting the synthesis weights (Fig.~\ref{fig:overview}(c)).
These weights can be thought of as the ``coefficients'' under the scattering ``bases'' learned by the expert DNNs in the DSN for linear synthesis.
To perform this task, we design a GTN that extracts multi-scale spatial features from the matching preprocessed hologram (see more details in Section~\ref{sec:preprocess}) and outputs three scalar numbers $\{\alpha_{1}, \alpha_{2}, \alpha_{3}\}$ that sum to unity:  $\sum_{i=1}^3 \alpha_{i} = 1$ (see more details in Section~\ref{sec:network} and Fig.~\ref{fig:s_GTN}).
In this way, the GTN performs a holistic analysis of the level of input scattering artifacts.
It enables adaptive mixture of  the features extracted by the expert encoders and linear synthesis of the expert decoder parameters. 
Therefore, the GTN can be considered as the primary agent for adaptability in the DSN.

\subsection{Training and Initialization Strategy} 
\label{sec:tr_init_strategy}

During the training, all the expert encoders and decoders in the DSN are co-trained with the GTN (see details in Section~\ref{sec:network}).
In this way, the GTN learns to optimally combine the expert encoders and decoders for varying levels of input scattering.
During the same time, the experts ($\tn{E}_{i}$ and $\tn{D}_{i}$) are observed to gain ``specializations'' to different scattering conditions, as discussed in detail later.
In order to impart broad adaptability to the DSN, the training data contains a diverse set of scattering levels (see details in Section~\ref{sec:dataset}).
\\

We investigate two DSN initialization strategies. 
In both cases, the GTN is initialized using the Xavier random weight scheme~\cite{glorot2010} while the parameters in the expert DNNs are initialized differently. 
In the first scheme, for which results are shown in the main text, each expert network in the DSN is initialized by first pre-training on a data set with a specific level of scattering density (see details in Section~\ref{sec:dataset}).
This is particularly helpful to illustrate the MoE type of working mechanism of the DSN, as we discuss in detail in Section~\ref{sec:GTN}.
In addition, instead of fixing the parameters of the pretrained experts for synthesizing the network, we continue the training process by co-training the expert networks with the random-initialized GTN to obtain the final DSN parameters.
\\

In the second initialization scheme, we remove the pre-training requirement and initialize all the experts and the GTN with Xavier random weights~\cite{glorot2010} (see details in Section~\ref{sec:network}).
Our results show that the DSN still converges to a multi-expert configuration despite different initialization methods (as shown in Fig.~\ref{fig:s_xavier_act}). 
We further demonstrate that regardless of the initialization scheme (i.e. with pre-trained or randomly initialized expert parameters), the DSN produces similar descattering performance, as shown in Fig.~\ref{fig:s_rvi}.
This highlights that the native DSN architecture can both impart specializations to its constituent experts, and enable adaptability by its inherent dynamic reconfiguration capability.
\\

\subsection{Performance in Seen Scattering Conditions in Simulation}
\label{sec:quant-sim}

\begin{figure}
	\centering
	\includegraphics[width=1\linewidth]{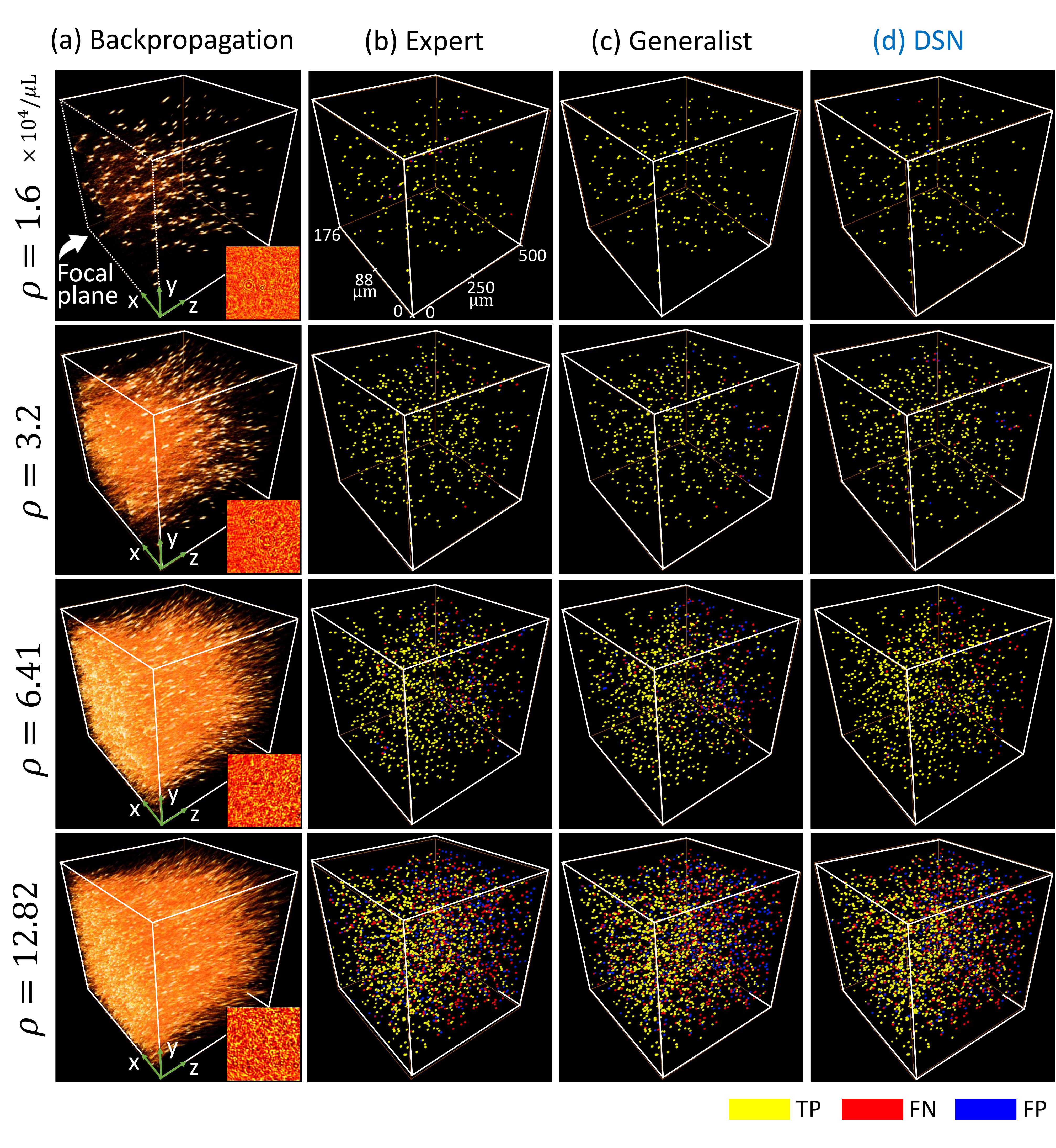}
	\caption{\textbf{Particle localization visualization.} 
		(a) 3D renderings of the backpropagated volumes demonstrate the scattering artifacts for various particle densities. The corresponding hologram is shown as the insets. (b-d) Particle 3D localization results are shown for the expert, generalist and DSN, respectively. 
		For low particle densities, all three DNNs perform similarly. 
		As the particle density increases, the DSN provides improved performance, measured by the true positives (TP, in yellow), false negatives (FN, in red), and false positives (FP, in blue). The numbers of TP, FP, and FN are quantified in Table~\ref{fig:s_count}, highlighting that the results from the DSN contain much fewer FPs at high particle density cases.}
	\label{fig:quali}
\end{figure}

First, we present 3D descattering performance of the DSN on ``seen'' scattering conditions (i.e. the scattering conditions used in the training).
For this purpose, we first simulate inline holograms from high-contrast (refractive index contrast $\Delta n$ = 0.26) micro-spheres (diameter $D$ = $1.0~\mu$m) randomly distributed in a 3D volume ($176.64\times 176.64\times 500~\mu$m$^3$). 
We generate data from four different particle densities ($\rho = \{1.6, 3.2, 6.41, 12.82\}\times 10^4$ particles/$\mu$L).
The density was chosen to increase by a factor of 2$\times$ in each step in order to cover a broad range of scattering levels, ranging from sparse to dense cases. 
To accurately model the multiple-scattering effects, we generate holograms using the beam-propagation method (BPM), whose high accuracy has been established experimentally by our recent work~\cite{wang2021}. 
Additional details about the simulation are provided in Section~\ref{sec:simulator}.
To perform 3D reconstruction from each hologram using the DNNs, we first perform 3D holographic backpropagation as a prepossessing step (see details in Section~\ref{sec:holo-prop}). 
The backpropagated volume is contaminated by scattering artifacts, whose severity depends on the particle density, size, refractive index contrast, and generally worsens as the depth increases~\cite{tahir2019}. 
The backpropagated volumes, ground truth volumes, and the corresponding simulated holograms are used for training each DNN (see details in Sections \ref{sec:preprocess} and \ref{sec:dataset}).
%
%
%
\\

Next, we apply each trained DNN to remove the 3D scattering artifacts in order to localize the particles.
For each density, we first visualize examples of 3D particle localization results of the DSN in Fig.~\ref{fig:quali} and compare them with two alternative strategies, including the expert network  and the generalist network at each matching scattering condition.
For this initial comparison, both the expert and the generalist networks use the same V-net architecture and the same number of trainable parameters, as detailed in Section~\ref{sec:network}, Fig.~\ref{fig:s_Vnet} and Table~\ref{fig:gen_parameters}.
In Section~\ref{sec:G3x}, we further benchmark the performance of a ``larger'' generalist network having approximately the same number of trainable parameters as the DSN and 3$\times$ numbers of parameters as the expert network.
The generalist network is trained using the same data set as the DSN.
%
%
To perform DNN testing, preprocessed holographically backpropagated 3D volumes are used as the input, which are computed from holograms never used during the training.
The input contains scattering artifacts that increase significantly with the particle density, as visually evident from Fig.~\ref{fig:quali}(a).
A patch from the corresponding hologram is also shown as the inset in Fig.~\ref{fig:quali}(a) to demonstrate how the fringe pattern qualitatively varies with the particle density and begins to resemble speckle patterns at high particle densities.
In Fig.~\ref{fig:quali}(b)--(d), we visualize the 3D particle localization results by labeling each particle in the network's output by true positive (TP), false positive (FP), or false negative (FN).
The labeling procedure is detailed in Section~\ref{sec:metrics}.
The numbers of TP, FP, and FN particles for each density in the test set (refractive index contrast = 0.26, particle size = $1~\mu$m, 10 volumes) are quantified in Table~\ref{fig:s_count}.
For relatively low particle densities $(\rho \le 3.2 \times 10^4~\tn{particles/}\mu\tn{L})$, all three DNNs demonstrate similar and highly accurate localization.
For higher particle densities $(\rho \ge 6.41  \times 10^4~\tn{particles/}\mu\tn{L})$, the DSN outperforms both the generalist and the matching expert.
\\

\begin{figure} [th]
	\centering
	\includegraphics[width=1\linewidth]{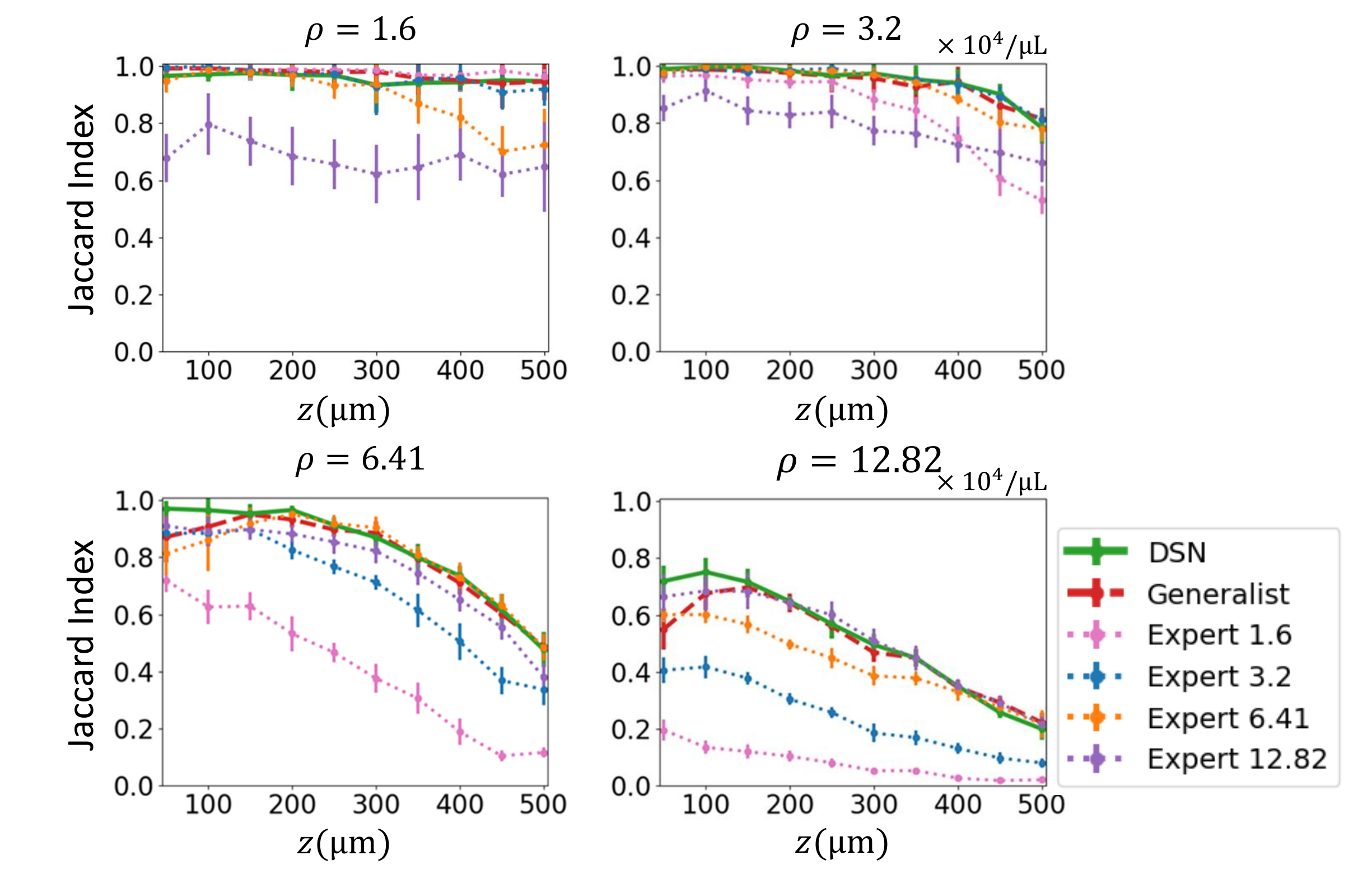}
	\caption{\textbf{Quantitative performance evaluation on seen densities.} 
		Particle localization performance is quantitatively compared between the DSN, the generalist, and the expert DNNs using JI. 
		Each plot indicates the results on a test data set at the density labeled above each plot and with particle diameter $1.0~\mu$m and refractive index contrast $0.26$. 
		``Expert~$\rho$'' represents the expert DNN trained on the data set with the density $\rho$ ($\times 10^{4}$ particles/$\mu$L).
		The DSN generally provides higher accuracy than both the generalist and the matching expert, in particular for high densities ($\rho \geq 6.41 \times 10^4$ particles/$\mu$L).
	}
	\label{fig:quant}
\end{figure}

A particular challenge of holographic 3D localization is that the scattering artifacts become more severe at greater depth due to the accumulation of coherent diffraction effects from shallower depths~\cite{tahir2019}. 
In addition, the initially estimated particles from the holographic 3D backpropagation are elongated more severely at greater depth due to decreasing effective numerical aperture (NA), which aggravates the ``missing-cone'' artifacts~\cite{wang2021}, as clearly seen in Fig.~\ref{fig:s_bprop}. 
These confounding factors imply that even for a fixed particle density, there exists a significant amount of variations in terms of signal fidelity (from the particle) and scattering noise in the input volume.
As a result, we expect depth-dependent reconstruction performance. 
We quantify the performance of the DSN by comparing the estimated and the ground-truth particle locations using the Jaccard Index (JI) similarity score across the depths. 
We evaluate the depth-wise localization performance by 
quantifying the statistics of the JI on every 10 axial slices (i.e. every $50~\mu$m) in the reconstructed 3D volumes. 
Additional details about the quantitative evaluation are provided in Section~\ref{sec:metrics}.
The same evaluation procedure is also applied to the reconstructions from the generalist and all the expert DNNs.
We use ``Expert~$\rho$'' to denote the expert DNN trained on the data with a specific particle density of $\rho$~$\times 10^{4}$~particles$/\mu$L.
The localization accuracy curves are shown in Fig.~\ref{fig:quant}, in which each plot quantifies the performance of all the DNNs on the same testing data set with the testing particle density labeled above each plot.
Each testing data set at a given particle density comprises of 10 volumes of particles with the same diameter (1.0~$\mu$m) and refractive index contrast (0.26) and different randomly generated 3D locations.
Each error bar quantifies the mean and the standard deviation of the JI computed on all the reconstructed particles within the corresponding 10 slices of each testing volume and across all the testing volumes. 
\\

We generally observe that the DSN outperforms the generalist and the matching expert DNN at high particle densities ($\rho\ge6.41\times 10^4$~particles$/\mu$L).
Each expert DNN generally performs well on the testing data with the matching  density; however, its performance degrades when the testing density is different from the training.
Interestingly, both Expert 3.2 and Expert 6.41 provide robust performance for the first three lower density cases ($\rho\le6.41\times 10^4$~particles$/\mu$L).
Expert 12.82 provides robust performance for the last two higher density cases ($\rho\ge6.41\times 10^4$~particles$/\mu$L).
These observations suggest that there indeed exist {\it generalizable multi-scale spatial features across different scattering densities}.
This heuristically provides the foundation to form and combine a set of scattering ``bases''  for dynamic synthesis in the DSN.
Combining all the scattering cases in a brute-force manner for training the generalist network can provide good performance across all the scattering levels, and achieve improved generalizability to different particle densities as compared to the expert DNNs. 
Such behavior has been observed in our previous works as well~\cite{li2018deep,li2021}.
However, the generalist generally provides lower accuracy than the expert DNN trained at the matching condition and the DSN.
\\

Additionally, we make the following observations for individual testing densities.
For ${\rho=1.6}\times 10^{4}$ particles$/\mu$L case, the DSN (green solid line) has good performance (JI~$>0.9$) across all depths, but is slightly worse than the expert (pink dotted line).
Upon visual inspection, we attribute this to the following observation. 
Due to the low particle density and the need to crop the full measurement to small patches, both the training and testing sets contain a small number of ``empty'' sub-volumes with no particles, i.e. the ground-truth volume is all zero.  
The DSN tends to ``hallucinate'' with FPs in these empty regions. 
The matching expert suffers less from this artifact possibly because its training data set contains approximately 4$\times$ more empty volumes than the DSN (see details in Section~\ref{sec:dataset}).
For ${\rho=3.2}\times 10^{4}$ particles$/\mu$L case, the DSN performs similarly to both the generalist (red dashed line) and Expert 3.2 (blue dotted line). 
Due to the increased density, all training and testing patches are not empty and hence do not suffer from the issues found in the lower density case. 
The performance begins to degrade as the particle density increases to $\rho=6.41\times 10^{4}$ particles$/\mu$L, especially at large depths ($z\ge400~\mu$m).
The DSN outperforms both the generalist (red dashed line) and Expert 6.41 (orange dotted line).
The accuracy of Expert 1.6 decreases sharply in this case, which shows that it cannot generalize and handle the increased scattering artifacts.
Expert~6.41 suffers from relatively lower accuracy at the first $200~\mu$m shallower depths. 
We attribute this to the significant variations in the signal-to-noise level throughout the $500~\mu$m imaging volume that arise due to the depth-dependent scattering artifacts at this relatively high density. 
As seen from the example shown in Fig.~\ref{fig:s_bprop}, at deeper regions, the particles are harder to be distinguished from the background scattering ``noise'' in the backpropagated volume.
Since the expert DNN is trained to handle these variations using a loss function that weighs the prediction error {\it equally} at all depths (see Section~\ref{sec:network}), the training process tends to find a balance between the high-noise deep regions and low-noise shallow regions, which results in worse performance at the shallow depths.
The same problem is also observed in the generalist when applied to $\rho= 6.41\times 10^{4}$ particles$/\mu$L and $\rho= 12.82\times 10^{4}$ particles$/\mu$L cases.
This issue is much alleviated by the DSN, likely because the DSN dynamically fuses the learned features during the inference, which can provide a better feature representation of the scattering input than the expert  and generalist networks.
To illustrate that different spatial features are extracted by the expert encoders in the DSN, we show example feature maps for two particle densities in Fig.~\ref{fig:feature_diff_scatter}.
\\

\subsection{DSN Generalizes to Unseen Scattering Conditions}
\begin{figure}
	\centering
	\includegraphics[width=0.91\linewidth]{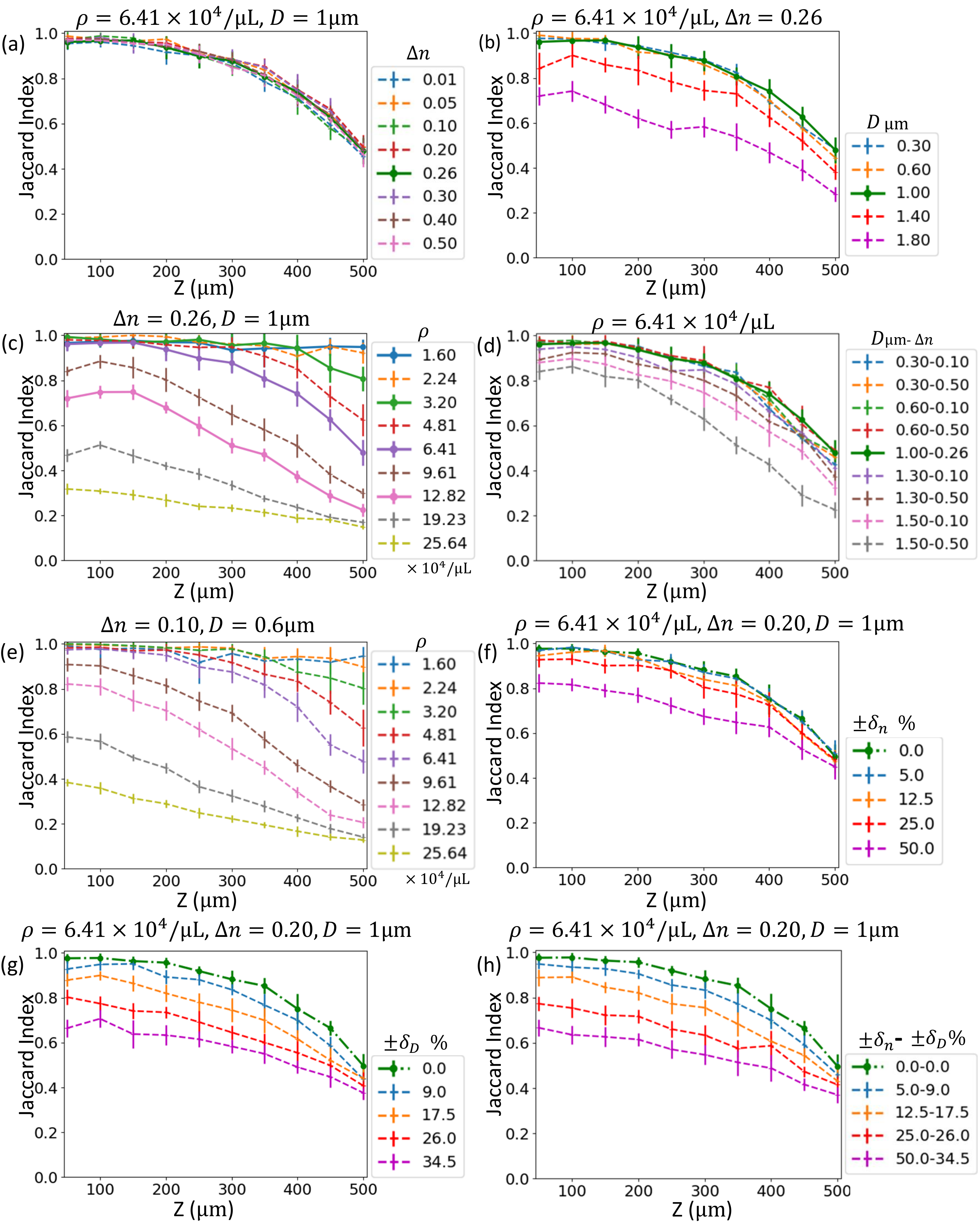}
	\caption{\textbf{Generalization of the DSN to unseen scattering conditions.} 
	The DSN demonstrates robust descattering for a continuum of scattering levels covering a wide range of refractive indices, particle sizes, and particle densities.
	The baseline seen cases are shown in solid lines; the unseen test conditions are in dashed lines. 
	The testing cases include:
	(a) unseen refractive index contrast ($\Delta n$);
	(b) unseen particle size ($D$); 
	(c) unseen particle density ($\rho$);
	(d) unseen refractive index contrast and particle size ($D-\Delta n$);
	(e) unseen refractive index contrast, particle size and density;
	(f) Uniformly distributed random refractive index contrast,  $\pm\delta_n$\% denotes the variation range with respect to the central refractive index contrast;
	(g) Uniformly distributed random particle size, $\pm\delta_D$\% denotes the variation range with respect to the central size; 
	(h) Uniformly distributed random refractive index and particle size ($\pm\delta_n-\pm\delta_D$). 
	In (f)-(h), the green dash-dotted line is the baseline unseen case at $\rho=6.41\times 10^{4}$~particles$/\mu$L, with a fixed refractive index contrast $\Delta n=0.20$ and a fixed particle size $D=1.0~\mu$m. 
		}
	\label{fig:unseen}
\end{figure}

Next, we demonstrate the DSN’s ability to generalize to ``unseen'' scattering conditions on testing data whose particle densities, sizes, and refractive index contrast were never used during the training.
To provide a holistic view of the DSN's descattering performance, we show the localization JI as a function of imaging depths for ``unseen" conditions (dashed lines) in Fig.~\ref{fig:unseen}, along with the JI curves for the ``seen'' conditions (solid lines, identical to the corresponding curves in Fig.~\ref{fig:quant}). 
The DSN is trained  using only the data from the seen cases with a fixed refractive index contrast 0.26, a fixed particle size 1.0~$\mu$m, and the four different particle densities, as detailed in Section~\ref{sec:dataset}. 
\\

The generalization power of the DSN is quantified in Fig.~\ref{fig:unseen}.
Our results demonstrate that the DSN adapts to a wide range of scattering conditions and provides robust descattering performance. 
First, we test the DSN's generalization by changing either the refractive index, the particle size, or the density. 
In Fig.~\ref{fig:unseen}(a), we show that the DSN is highly robust to the change of the refractive index contrast ($\Delta n$), yielding negligible localization performance variations for the entire tested $\Delta n$ range (from 0.01 to 0.50). 
A possible explanation is that our data normalization procedure (see Section~\ref{sec:preprocess}) facilitates the DSN to focus on scattering-dependent spatial features by removing the ``superficial'' changes in the low-order (including mean and standard deviation) intensity statistics (see Fig.~\ref{fig:preprocessing}), and thus provides superior generalization performance to the refractive index contrast.
In Fig.~\ref{fig:unseen}(b), we show that the DSN is also robust to the change of the particle size, especially when the testing size is smaller than the trained size. 
The performance degrades to JI~$<$~0.8 only when the particle size is increased by $80$\%.  
This is expected because the increased particle size results in novel spatial features that are not learned by the DSN that is trained using a fixed particle size.
In Fig.~\ref{fig:unseen}(c), we quantify the DSN's performance at various unseen particle densities.  
The DSN demonstrates a smooth transition between different scattering levels, regardless of whether they are seen or unseen. 
This implies that the overall decrease in performance at higher particle densities is more associated with the increased level of scattering artifacts in the input volume (as expected), but less affected by whether the scattering level has been used during the training (i.e. network generalization). 
\\

Next, we show the generalization of the DSN on data sets with both unseen particle size and unseen refractive index at a given density in Fig.~\ref{fig:unseen}(d).  
Similar to the observations in Fig.~\ref{fig:unseen}(a)--(b), the DSN  is still robust to these changes.
We further test the DSN under a fixed unseen particle size and a fixed unseen refractive index for a variety of particle densities in Fig.~\ref{fig:unseen}(e).
By comparing the results in Fig.~\ref{fig:unseen}(c) and (e) (which have matching particle densities), we confirm that the DSN's descattering performance is indeed primarily controlled by the particle density.
\\

Finally, we test the DSN on volumes containing particles with randomly distributed refractive indices and sizes.
To quantify the effect, we assume both the refractive index and the particle size follow uniform distributions (see details in Section~\ref{sec:dataset}). 
The performance of DSN gradually degrades as the variation range increases.  
In Fig.~\ref{fig:unseen}(f), the results show that the DSN is robust to random variations of the refractive index up to around 50$\%$ amount of fluctuations, beyond which the performance degrades to JI~$<$~0.8 in most of the depths.
The fact that the DSN is robust to the global change in the refractive index contrast (Fig.~\ref{fig:unseen}(a)) but degrades on randomly distributed values (Fig.~\ref{fig:unseen}(f)) suggests that the network's performance is limited by the dynamic range in the measurement. 
Low-contrast particles result in weaker in-focus signals in the backpropagated volume and thus are harder to reconstruct in particular in the presence of high-contrast particles, both of which are surrounded by the same strong background scattering noise.
In Fig.~\ref{fig:unseen}(g), the random variations of the particle size result in reduced accuracy, similar to those observed in Fig.~\ref{fig:unseen}(b). 
The results indicate that the DSN is robust to random variations of the particle size up to around 26$\%$ amount of fluctuations, beyond which the performance degrades to JI~$<$~0.8 in most of the depths.
In Fig.~\ref{fig:unseen}(h), random variations in both the refractive index and the particle size also lead to smoothly reduced performance.
By comparing Figs.~\ref{fig:unseen}(g) and (h) (without and with particle size variations), we conclude that the generalization of the DSN is primarily affected by the particle size variations.
\\

Overall, these results highlight that the DSN can perform adaptive 3D descattering for a continuum of scattering levels beyond the three specific conditions represented by the experts in its native architecture and the four conditions used in the training data set.
This bypasses the need for training a large bank of expert DNNs to handle data with a wide range of scattering conditions. 
The performance of the DSN can potentially be improved, particularly at high particle densities, by increasing the number of expert networks within the DSN, based on our study in Fig.~\ref{fig:s_2v3}.
The performance can also be improved by diversifying the training data to include additional particle densities, sizes, and refractive indices to refine the dynamic synthesis with improved generalization in the future.

\subsection{Comparison of the DSN and Generalist}
\label{sec:G3x}
\begin{figure}
	\centering
	\includegraphics[width=0.59\linewidth]{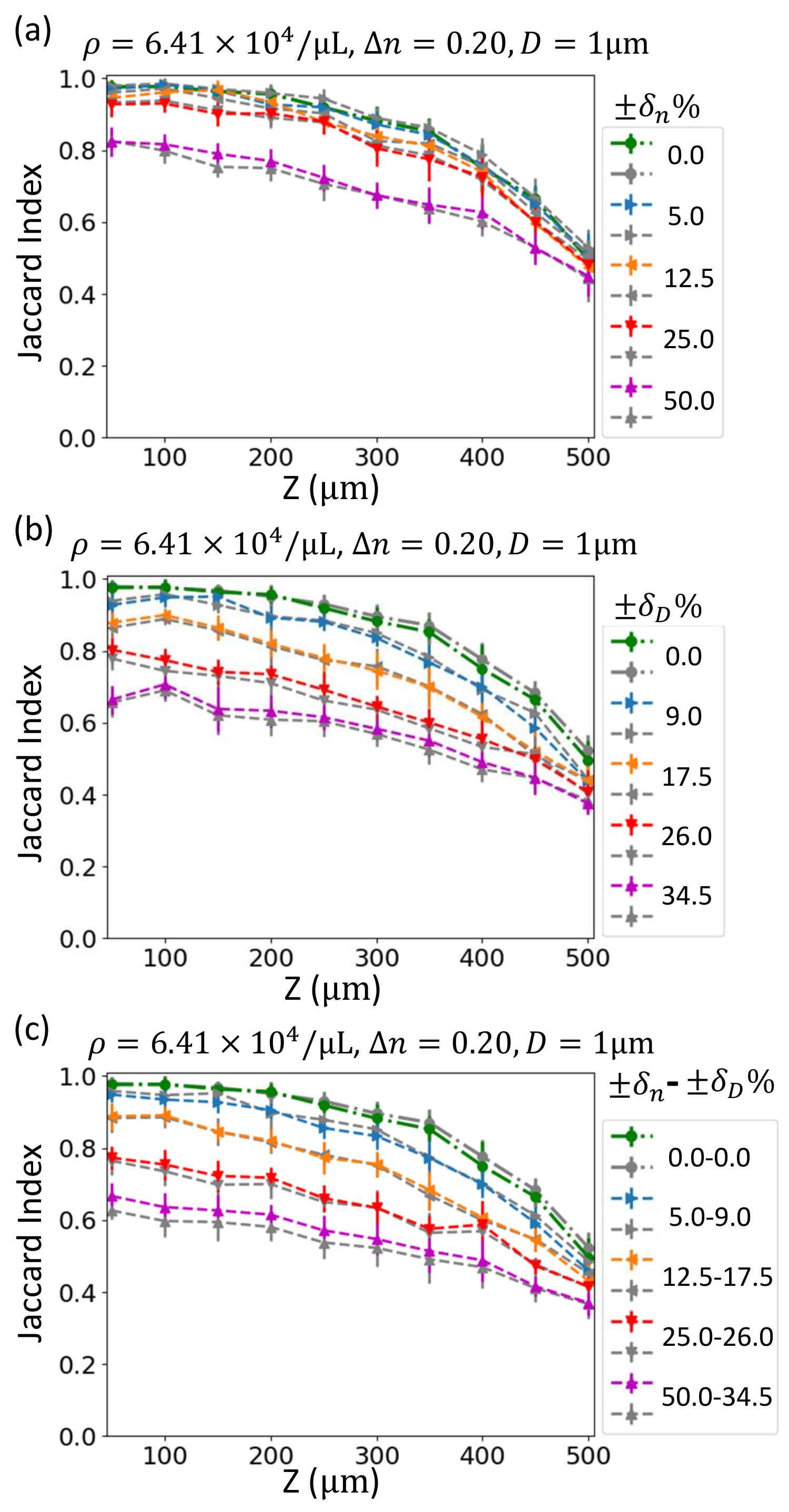}
	\caption{\textbf{Comparison of the generalization capability between the DSN and the $3\times$ generalist.} 
	The testing volumes contain particles with (a) uniformly distributed random refractive index contrast; (b) uniformly distributed random particle size;
	and (c) uniformly distributed random refractive index contrast and particle size.
	The testing particle density ($\rho$), central refractive index ($\Delta n$), and central particle size ($D$) are shown in the title in each plot.
	The variation ranges of the refractive indices and the particle sizes are marked by $\pm\delta_n\%$ and $\pm\delta_D\%$, respectively.
	The green dash-dotted line is the baseline unseen case at $\rho=6.41\times 10^{4}$~particles$/\mu$L, with a fixed refractive index contrast $\Delta n=0.20$ and a fixed particle size $D=1.0~\mu$m. 
	The DSN results are shown in dashed lines in different colors for different cases; the corresponding 3$\times$ generalist results are shown in gray dashed lines with the matching markers.
	The DSN outperforms the 3$\times$ generalist as the scattering condition become more different from the baseline case.}
	\label{fig:DSN_3X_gen_random}
\end{figure}

We further analyze and compare the DSN with two generalist networks (trained on the same data set as the DSN). 
As described in Section~\ref{DSN_explaination} and shown in Fig.~\ref{fig:feature}(a), the DSN dynamically combines features extracted by different expert networks to synthesize an optimal feature representation of the input.
This means that the synthesized DSN during inference uses the {\it same} number of feature maps as the expert, as illustrated in Fig.~\ref{fig:feature}(b).
The DSN achieves this by optimally utilizing the bases learned by the expert networks through a linear feature synthesis.
%
%
As a comparison, an alternative ``brute-force'' method to increase the expressing power of a neural network is to directly increase the number of parameters (and hence features), by increasing the number of channels in the convolutional layers and/or increasing the number of layers in the network.
Since the number of trainable parameters in an untrained DSN is approximately 3$\times$ the number in the expert, we also compare the DSN with a ``larger'' generalist network containing the same number of trainable parameters, termed ``$3\times$ Generalist''.
The hyper-parameters in the $3\times$ Generalist is heuristically optimized.
It contains the same number of layers as the expert while having a larger number of convolutional channels in each layer, as detailed in Section~\ref{sec:Gnetwork} and Table~\ref{fig:3_gen_parameters}. 
\\

First, we compare the DSN with the baseline generalist having the same number of feature maps and show that the DSN optimized feature maps indeed result in improved descattering performance.
%
%
In Section~\ref{sec:quant-sim}, we have shown that the DSN outperforms the baseline generalist on seen cases (Fig.~\ref{fig:quant}).
Next, we compare their generalization capability on unseen scattering conditions.
The localization performance of the baseline generalist on the cases in Fig.~\ref{fig:unseen} (shown for the DSN) are provided in Fig.~\ref{fig:unseen_1Xgen}.
A direct comparison at different unseen densities are also plotted on the same figures in Fig.~\ref{fig:s_unseen_1X}.
%
It is clearly evident that the DSN performs better than the baseline generalist under all different unseen cases.
%
%
%
\\

Next, we compare the DSN with the $3\times$ generalist.
First, we compare their performance on seen scattering conditions in Fig.~\ref{fig:s_seen_3X}.
The DSN and the $3\times$ generalist have similar performance for the first three lower densities ($\rho = \{ 1.6, 3.2, 6.41\}\times10^4$ particles$/\mu$L).
The $3\times$ generalist slightly outperforms the DSN at the highest seen density case ($\rho=12.82\times10^4$ particles$/\mu$L), which we attribute to its higher expressing power.
Next, we compare the two networks' generalization capability on unseen scattering conditions.
Some representative direct comparisons on the most challenging cases containing randomly distributed particle sizes and refractive indices are shown in Fig.~\ref{fig:DSN_3X_gen_random}.
Additional results of the $3\times$ generalist for the scattering conditions matching those in Fig.~\ref{fig:unseen} are shown in Fig.~\ref{fig:unseen_3Xgen}.
The DSN generally performs better than the $3\times$ generalist when the testing condition deviates from the baseline case. 
These results imply that by increasing the number of feature maps in the $3\times$ generalist, it allows the network to learn more information that is particularly beneficial for descattering the seen high-density cases.
However, the larger network also tends to overfit the training data, which results in degraded performance in unseen cases. 
While it is possible to alleviate overfitting by using a larger training data set, we use the same training set as the DSN for fair comparison.
Since the DSN still maintains the same feature map number as the  baseline generalist network, it is less prone to overfitting using the same training data set.
In addition, the $3\times$ generalist requires higher computational cost than the DSN during inference (more network parameters), as detailed in Sections~\ref{DSN_explaination}, ~\ref{sec:network} and \ref{sec:Gnetwork}, which makes the $3\times$ generalist approach less appealing. 
\\

\begin{figure}
	\centering
	\includegraphics[width=1\linewidth]{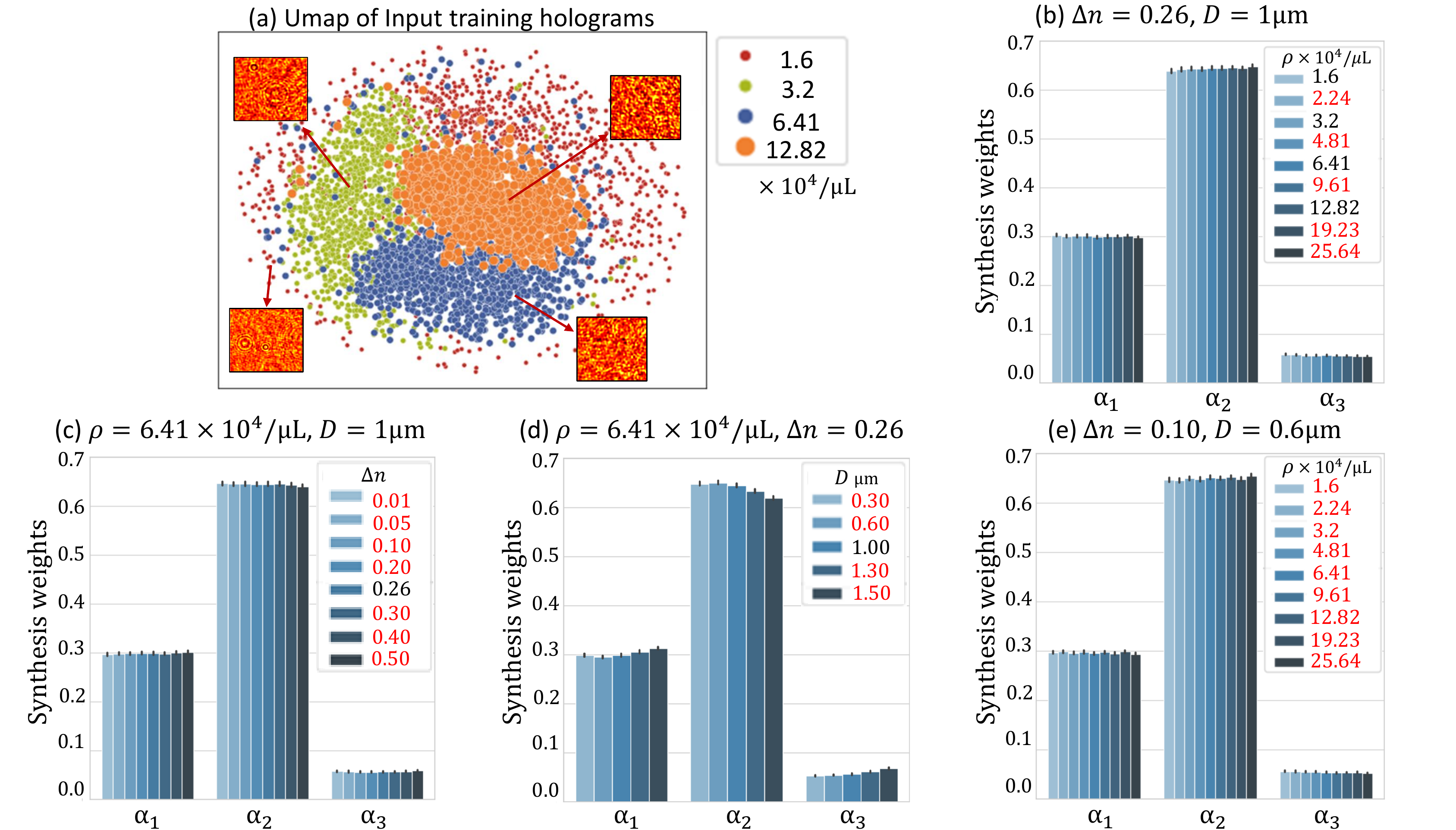}
	\caption{\textbf{Gating network analysis.} 
		(a) The 2D UMAP representation for the measured hologram patches used in the training. Hologram patches with similar particle densities cluster together, indicating their statistical similarity. Thus, the hologram can serve as a proxy to the scattering level and is used as the input to the GTN to predict the synthesis weight. 
		The synthesis weights are shown for various seen (legend in black) and unseen (legend in red) scattering conditions, including (b) different particle densities, (c) different refractive indices, (d) different particle diameters, and (e) different particle densities with an unseen particle refractive index and size.
		The synthesis weights are consistent for each condition and further tailored to each input, as quantified by the mean and standard deviation for each case. 
		The larger values of $\alpha_2$ indicate the major contributions of $\tb{F}_{2}$ and $\tn{D}_2$ to the DSN. 
		As the particle density increases, $\alpha_{2}$ increases while $\alpha_{1}$ and $\alpha_{3}$ decrease regardless of whether the particle size and refractive index are seen (b) or unseen (e).
		}
	\label{fig:umap}
\end{figure}

\subsection{Analysis of Gating Network}
\label{sec:GTN}

To elucidate on the working mechanism of the DSN, next we discuss how the GTN dynamically synthesizes the feature maps and network parameters, and in what manner specializations are imparted to the experts within the DSN.
\\

The GTN enables the adaptability of the DSN by generating the synthesis weights for combining the feature maps learned by the expert encoders and parameters of the expert decoders based on the ``feedback signal'' from the input hologram. 
It is built on the idea that the raw hologram contains sufficient information to infer the underlying scattering condition.
To test this, we use the state-of-the-art unsupervised dimensionality reduction technique, UMAP~\cite{mcinnes2018} to visualize the joint distribution of 5120 hologram patches in the training set distributed equally across the four particle densities in Fig.~\ref{fig:umap}(a) (see more details in Section~\ref{sec:UMAP}).
For this, we follow the computational procedure established in our recent work~\cite{li2021}.
As shown in Fig.~\ref{fig:umap}(a), the preprocessed hologram patches with similar underlying particle densities cluster together in the low dimensional 2D UMAP space. 
This indicates that the holograms contain intrinsic spatial features to inform the underlying scattering level.
This information is extracted by the GTN to adaptively set the synthesis weights for descattering the input volume.
\\

After co-training with the experts within the DSN, the task of the GTN is to {\it reliably} predict the synthesis weights.
A necessary condition is that the GTN should generate similar synthesis weights for holograms measured from similar scattering conditions.
To test this, we gather the statistics of the predicted synthesis weights and compute the mean and the standard deviation of each weight $\alpha_i$ for several different scattering conditions, including a wide range of seen (legend in black) and unseen (legend in red) particle densities in Fig.~\ref{fig:umap}(b), refractive indices in Fig.~\ref{fig:umap}(c), particle sizes in Fig.~\ref{fig:umap}(d), and different particle densities at an unseen refractive index ($\Delta n=0.10$) and an unseen particle size ($D=0.6~\mu$m). 
The results show that the standard deviation is much smaller than the mean for all the cases studied, indicating the consistency of the GTN-predicted synthesis weight. 
The small weight variations also indicate that the GTN does {\it not} perform a ``rigid'' classification of the scattering condition, but rather predicts a set of synthesis weight in a continuous space. 
It automatically chooses a different linear mixture of experts (i.e. feature maps and decoder parameters) when the scattering condition changes.
Thus, by engaging a unique combination of experts for each input, the DSN can not only adapt to different scattering levels but also provide a fine-tuned synthesized network to each input.
\\

Next, we illustrate the different specializations of the experts in a trained DSN.
Since the synthesis weights determine the activation of each expert in the synthesized network, we argue that the specialization of each expert can be qualitatively inferred by the statistics of the synthesis weights.
A similar interpretation was recently provided  by Yang {\it et al.}~\cite{yang2020multi} for analyzing a dynamically synthesized network for robotic locomotion. 
To demonstrate the expert specializations, we plot the synthesis weights, $\alpha_{i \in \{1,2,3\}}$, for a variety of particle densities (Fig.~\ref{fig:umap}(b) and (e)), refractive index contrasts (Fig.~\ref{fig:umap}(c)), and particle sizes (Fig.~\ref{fig:umap}(d)).
We note that the overall contribution of $\alpha_{2}$ is always the largest compared to $\alpha_{1}$ and $\alpha_{3}$, making $\tb{F}_{2}$ and $\tn{D}_{2}$ the major contributors of any reconstruction regardless of the input.
As the particle density is varied from low to high, $\alpha_{2}$ increases while $\alpha_{1}$ and $\alpha_{3}$ decrease.
This trend is consistent regardless of whether the particle size and refractive index are seen (Fig.~\ref{fig:umap}(b)) or unseen (Fig.~\ref{fig:umap}(e)).
This indicates that the GTN adaptively utilizes more features from $\tb{F}_{2}$ to handle the increased scattering artifacts from the increased particle density.
As the refractive index contrast increases for a fixed particle density, the synthesis weights change only slightly (Fig.~\ref{fig:umap}(c)). 
Although the increased refractive index results in a greater amount of scattering from individual particles, the data normalization scheme (see Section~\ref{sec:preprocess}) removes any resulting variations in the low-order statistics (including mean and standard deviation).
As a result, the GTN is forced to pay more attention to spatial features induced by inter-particle scattering, which are expected to be less prominent as compared to changing the particle densities.
On the other hand, as the particle size increases for a fixed particle density, the changes in the spatial features result in decrease in
$\alpha_{2}$ and increase in $\alpha_{1}$ and $\alpha_{3}$. 
For all the tested data, $\alpha_{3}$ remains a small value, indicating that the corresponding expert feature maps $\tb{F}_{3}$ (and the expert decoder $\tn{D}_{3}$) only provide fine-tuned contributions.
To ensure that $\tb{F}_{3}$ and $\tn{D}_{3}$ are not redundant, we trained a DSN with only two pairs of expert encoders and decoders in Fig.~\ref{fig:s_2v3}. 
The two-expert DSN under-performs the three-expert DSN, which indicates the importance of $\tb{F}_{3}$ and $\tn{D}_{3}$.
Overall, a smooth transition is observed for each synthesis weight $\alpha_{i}$ with respect to the input particle density, refractive index contrast, and particle size regardless of whether they are seen or unseen.
This corroborates with our earlier observation of a smooth transition in the DSN performance as the input varies across different scattering conditions.
\\

For the results presented so far, the DSN was trained using the pre-training initialization scheme (see Section~\ref{sec:tr_init_strategy}).
Specifically, $\tn{E}_{1}$($\tn{D}_{1}$), $\tn{E}_{2}$($\tn{D}_{2}$), $\tn{E}_{3}$($\tn{D}_{3}$) are first individually pre-trained on a specific particle density from $3.2, 6.41,$ and $12.82 \times 10^4$ particles$/\mu$L, respectively (see more details in Section~\ref{sec:dataset}).
We applied this procedure since it is useful to qualitatively associate the specializations of each expert with the particle density used in the pre-training, e.g. $\tn{E}_{1}$($\tn{D}_{1}$) is specialized in low particle density.
However, this leads to a conjecture that the expert specializations in the trained DSN shown in Fig.~\ref{fig:umap} is merely a consequence of the initial bias introduced by the pre-training.
While this might be the case, and even desirable for certain applications~\cite{yang2020multi}, we further demonstrate that even with the random initialization, the DSN converges to a state with similar expert specializations.
When we perform a similar synthesis weight analysis for the DSN trained using the random initialization scheme in Fig.~\ref{fig:s_xavier_act}, we can draw a similar conclusion based on our earlier discussions in this section.
We note that although different initialization results in different synthesis weight distributions due to the stochastic training process, the observed trend for each synthesis weight remains the same and the trained DSNs produce very similar descattering performance (Fig.~\ref{fig:s_rvi}). 
Overall, these results again highlight that the DSN architecture natively imparts specializations to its constituent experts, and enables adaptive descattering by dynamically synthesizing the feature maps and adjusting its network parameters.
\\
%

\begin{figure}
	\centering
	\includegraphics[width=0.8\linewidth]{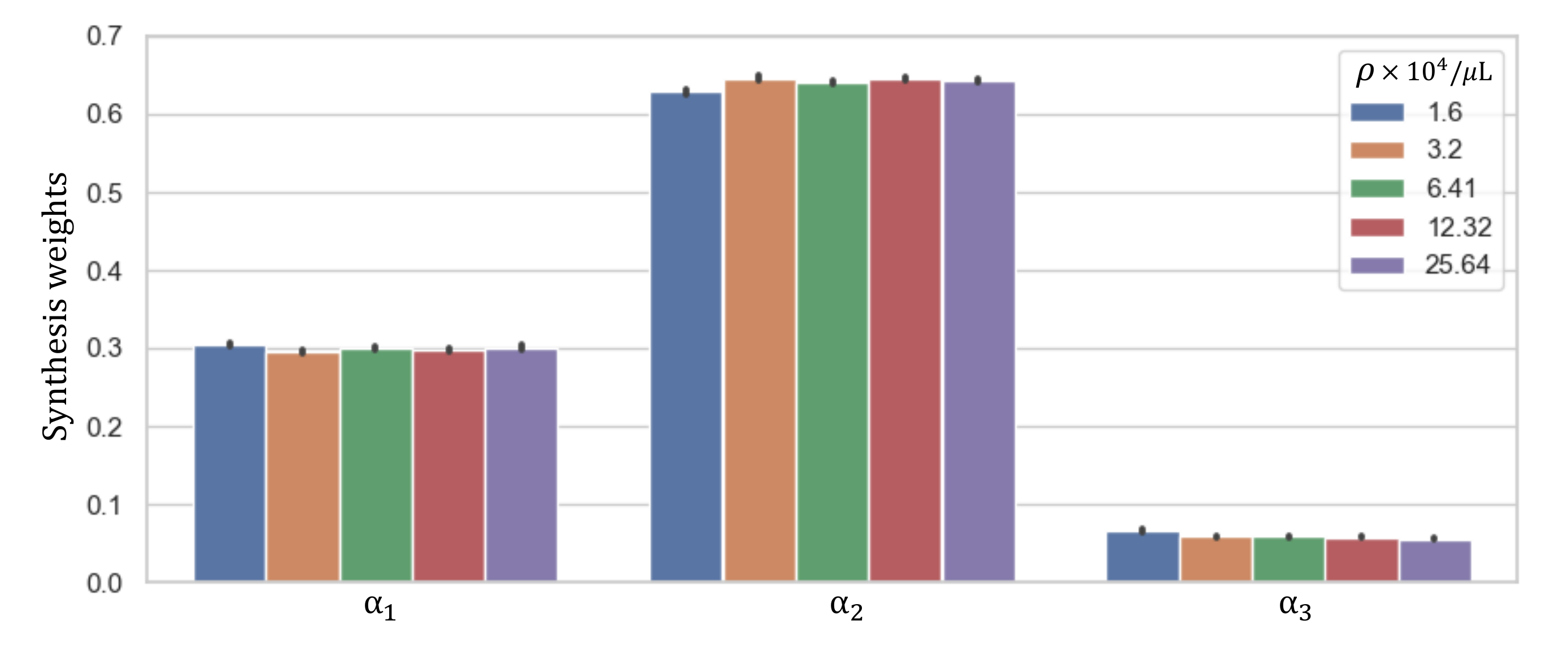}
	\caption{\textbf{Synthesis weights of the simulator-trained DSN for experimental data.} 
		The synthesis weights computed for the experimentally measured holograms match well with the corresponding simulation results. 
		The mean and standard deviation of each synthesis weight are calculated from 640 non-overlapping hologram patches from 10 experimentally measured holograms for each particle density.
		This analysis shows that the simulator-trained DSN can robustly adapt to experimental data.
		}
	\label{fig:exp_gtn}
\end{figure}

\begin{figure}
	\centering
	\includegraphics[width=0.98\linewidth]{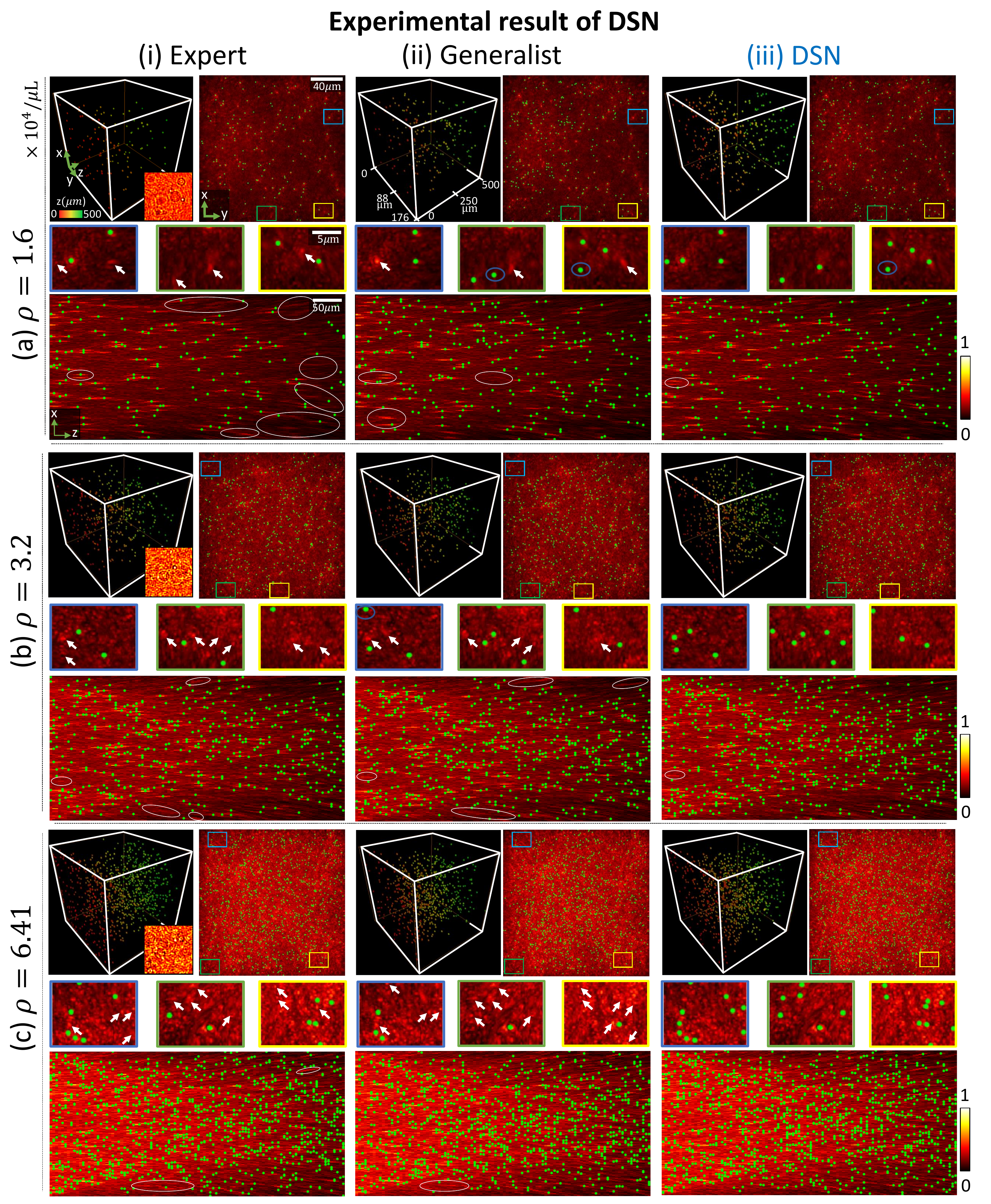}
	\caption{\textbf{Experimental results from simulator-trained DNNs.} 
    	Particle 3D localization is shown for experimentally measured holograms using the simulator-trained (i) expert, (ii) generalist, and (iii) DSN networks at three particle densities. 
    	Each panel shows (Top left) the 3D rendering of the localization result with depth color-coded particles, with an inset showing a zoom-in of the measured hologram, (Top right) the maximum intensity $z$-projection along with three zoom-in regions, and (Bottom) the $y$-projection of the DSN's 3D localization result (in green), overlaid on the respective $y$- and $z$-projections of the corresponding holographic backpropagations. 
    	Visually identified mis-detection regions are marked by white ovals in the $y$-projections and white arrows in the $z$-projections.
		}
	\label{fig:exp}
\end{figure}

\subsection{Generalization of Simulator-Trained DSN to Experimental Data}
Next, we assess the capability of the DSN on experimental measurements.
We capture in-line holograms of 3D samples consisting of polystyrene microspheres freely suspended in water, illuminated by a plane wave (more details in Section~\ref{sec:setup}).
We use five particle densities, including $\rho = \{1.6, 3.2, 6.41, 12.82, 25.64\}\times 10^4$ particles/$\mu$L, which increases by a factor of 2$\times$ in each step to cover a broad range of scattering levels. 
Following the same procedure as the simulation, each hologram is first back-propagated, preprocessed, and then input to the DSN to perform 3D descattering.
\\

Our results highlight that the {\it simulator-trained} DSN can be directly used in experiments and robustly perform particle 3D reconstruction across a wide range of densities.
We attribute the DSN's  generalization capability to two main factors.
First, the BPM model we employ to generate the training data in simulation can accurately model the multiple scattering process (more details in Section~\ref{sec:simulator}). 
The statistical properties of the simulated holograms closely match their experimentally measured counterparts, as quantified in our recent work~\cite{wang2021}.
This means that the input to the DSN does not suffer from ``domain shifts'', which bypasses the need for transfer learning~\cite{tan2018survey} or domain adaptation~\cite{ganin2016}, thus fulfilling a requirement for the simulator-trained DSN to generalize to experimental data, as also shown in our recent work~\cite{matlock2021p}.
Second, the DSN provides good generalization capability to different scattering conditions, which makes it robust to the sample variations present in real experiments, such as particle densities, particle sizes, beam imperfections, and different sources of noise.
To demonstrate this, we first plot the synthesis weights of the DSN for experimental data with five different particle densities in Fig.~\ref{fig:exp_gtn}.
Remarkably, the GTN reliably adjusts the synthesis weights for the experimental measurements in a manner that closely follows the behavior in the simulation (Fig.~\ref{fig:umap}(b)).
The value of each synthesis weight $\alpha_{i \in \{1,2,3\}}$ in the experiment matches well to the simulation at all particle densities. 
Small variations of the synthesis weights are observed for each particle density, as quantified by the standard deviation on each bar plot in Fig.~\ref{fig:exp_gtn}. 
This shows that the synthesis weights dynamically generated by the GTN are tailored to each scattering condition in the experimental measurements.
Similar to the simulation, $\alpha_2$ has the largest contribution compared to $\alpha_1$ and $\alpha_3$ across all the experimentally tested densities.
As the particle density increases, $\alpha_2$ generally increases, while $\alpha_1 $ and $\alpha_3$ decrease to adapt to the increased particle density.
\\

Next, we compare the localization results from the DSN, the expert, and the generalist networks on the experimentally measured holograms. 
The results on three representative cases with increasing densities ($\rho = \{1.6, 3.2, 6.41\}\times 10^4$ particles/$\mu$L) are shown in Fig.~\ref{fig:exp}; two additional cases at higher densities ($\rho = \{12.82, 25.64\}\times 10^4$ particles/$\mu$L) are shown in Fig.~\ref{fig:s_exp}.
To perform this comparison, both the expert for a given particle density and the generalist are also trained using only simulated data.
%
%
For each case, we show the depth color-coded 3D rendering of the particle localization, and the maximum intensity $y$- and $z$-projections overlaid on the corresponding projections of the 3D holographic backpropagation.
Although we do not have the ground-truth particle locations for the experiments, the amplitude distribution of the 3D holographic backpropagation can provide visual cues for identifying the particle positions~\cite{tian2010q}, especially at low particle densities.
For the lowest density case in Fig.~\ref{fig:exp}(a), the expert DNN visibly suffers from mis-detections especially at deeper depths, as highlighted by the white ovals in the $y$- and $z$-projections.
This is likely because the simulator-trained expert does not have sufficient generalizability to handle the unaccounted-for experimental variations.
The DSN and the generalist demonstrate more consistent localization results across the depths.
We attribute their superior generalization capability, as compared to the expert, to the increased feature representation power.
For the higher density cases in Figs.~\ref{fig:exp}(b), \ref{fig:exp}(c), and \ref{fig:s_exp}, it becomes challenging to establish the performance for every particle due to the increased scattering artifacts. 
However, we identify fewer clearly mis-detected regions for the DSN in the $y$- and $z$-projections.
Based on these visual inspections, we conclude that our experimental results are consistent with our simulation.
The simulator-trained DSN demonstrates robust 3D descattering, and provides improved 3D localization as compared to the expert and the generalist networks, in particular at higher densities and deeper depths.
\\

\section{Conclusion and Discussion}

In this study, we have presented and experimentally demonstrated a novel {\it adaptive} deep learning framework, termed dynamic synthesis network (DSN), for removing volumetric scattering artifacts.
We demonstrated the DSN's generalization capability of adaptively removing 3D scattering artifacts and achieving state-of-the-art performance for a wide range of scattering conditions. 
Broadly, the DSN provides a new ``mixture of experts'' network architecture, which 
adjusts the feature maps and network parameters ``on-the-fly'', to achieve adaptation to different input.
Our study also highlighted the utility of multiple-scattering simulator-based training that can enable generalization to real experimental data. 
This is particularly attractive since the paired ground-truth labels are hard to obtain in many scientific and biomedical imaging applications.
In our recent work, we have shown that this simulator-based training strategy can be applied to 3D quantitative phase imaging and achieve high-resolution reconstruction on complex biological samples~\cite{matlock2021p}.
We expect the synergy between physics-based large-scale simulation and data-driven models can significantly advance scientific/biomedical deep learning techniques.
Our proof-of-concept demonstration of the DSN focuses on holographic particle 3D imaging, which can find immediate applications in dynamic flow measurements~\cite{tian2010q, hinsch2002holographic, chen2021snapshot}, imaging cytometry~\cite{cheong2009flow, seo2009lensfree, merola2017tomographic}, and biological sample characterization~\cite{moon2009automated, su2012high}.
Broadly, we believe this new adaptive deep learning framework can be further adapted to many other imaging applications, including image denoising~\cite{weigert_content-aware_2018},  non-line-of-sight imaging~\cite{faccio2020non}, deep imaging in scattering tissue~\cite{badon2016smart,kang2015imaging}, and computational fluorescence microscopy~\cite{pegard2016compressive,xue2020single}.
\\

The DSN structure can be further improved in several aspects in the future.  
First, using the V-net as the ``backbone'' to construct the DSN for processing volumetric data poses challenges in scaling up the number of experts in the DSN due to the large computational cost.
However, increasing and thus diversifying the expertise is fundamental to push the performance limit of the DSN as shown in this study.
Future work may investigate lightweight network structures for processing volumetric data~\cite{matlock2021p, huang2020, kang2021} to reduce the computational cost. 
Second, in terms of the reconstruction accuracy, the DSN is still limited by its single-scattering approximant backpropagation input, which becomes particularly limiting at high particle densities.
Our recent work has shown that the model-based reconstruction based on the BPM can significantly improve reconstruction quality at strong scattering conditions and large imaging depths~\cite{wang2021}.
Thus, a promising future direction is to embed multiple-scattering physics into the DSN framework to further push the descattering performance~\cite{goy2019high}.
Third, the DSN's generalization capability can be  improved by further diversifying the training set to include different particle size and refractive index variations. 
A challenge to handle scattering samples with large contrast variations is the limited dynamic range in the measurement that makes the recovery of low-contrast scatterers unstable~\cite{matlock2021p,goy2019high}. 
Novel data augmentation techniques may be explored to overcome this issue~\cite{matlock2021p}.
Finally, the feature synthesis process in the DSN is similar to that used in the ``attention'' mechanism~\cite{8447284}. 
Thus, it is possible to refine the feature synthesis scheme by using novel attention learning architectures. 
%

\section{Materials and Methods}
\label{sec:method}
\subsection{DSN network design and implementation}
\label{sec:network}
The expert and baseline generalist networks have the same network architecture (shown in Fig.~\ref{fig:s_Vnet}) that is based on the V-net framework~\cite{milletari2016}.
It consists of an encoder and decoder with skip-connections for high-resolution feature forwarding.
The encoders and decoders in the DSN (shown in Fig.~\ref{fig:s_exp_struc}) are also based on the same V-net structure, since they are derived from the expert DNNs.
The GTN within the DSN is a relatively simple DNN following the VGG structure~\cite{simonyan2014} to predict the synthesis weights $\alpha_{i}$. 
The detailed implementation of the GTN is shown in Fig.~\ref{fig:s_GTN}.
The synthesis weights $\alpha_{i}$ are used to adaptively mix the extracted feature maps from the three expert encoders and to synthesize the parameters of the synthesized decoder in the DSN, as detailed in Fig.~\ref{fig:overview}.
\\

The DSN framework adaptively combines multiple expert DNNs to achieve dynamic synthesis. 
In the first stage, three expert encoders $\tn{E}_{i}$ independently perform multi-scale feature extractions from the preprocessed holographically backpropagated volume, where $i \in \{1,2,3\}$ represents the expert index.
We use $\tb{F}_{i}$ to represent all the extracted multi-scale feature maps from these expert encoders within the DSN, which are then used to compute the dynamically synthesized features $\tb{F}_{s}$ by the following linear combination, $\tb{F}_{s} = \sum_{i} \alpha_{i} \tb{F}_{i}$.
These synthesized features at the corresponding spatial scales are {\it directly}  passed to the synthesized decoder by the skip connections {\it without} any additional encoding procedures.
The synthesized ``latent'' feature maps (code) at the ``bottleneck'' (having the lowest dimension $8\times8\times7$) are passed to the synthesized decoder.
Effectively, the latent feature maps represent the adaptively optimized feature representation of the input under the three bases learned by the experts within the DSN.
This dynamic synthesis of the encoder is illustrated in Fig.~\ref{fig:s_exp_struc}. 
\\

On the decoding end, the DSN decoder is dynamically synthesized from the multiple expert decoders by $\tn{D}_{s} = \sum_{i} \alpha_{i} \tn{D}_{i}$, where $\tn{D}_{s}$ represents the network parameters of the synthesized decoder, and $\tn{D}_{i}$ represents the expert decoder parameters within the DSN.
Every network parameter in the synthesized decoder is thus a linear combination of the corresponding expert decoder parameters, whose proportions are determined by the GTN.
More details about the dynamic synthesis of the decoder are provided in Fig.~\ref{fig:s_exp_struc}(b). 
\\

To formulate the training loss function, we treat the ground truth volume as binary-valued, in which the sample contains discrete particles $\{1\}$ separated by the background $\{0\}$.
Accordingly, we employ the binary cross entropy as the loss function, which has shown to promote sparsity in the reconstruction~\cite{li2018deep}. 
The loss $\mathcal{L}(\tb{W})$ is defined by
\begin{equation}
	(\mathbf{W}) = 
	- \frac{1}{N} \left(\sum_{i=1}^{N} y_i \log \textrm{P}(y_i = 1 | \mathbf{X};\mathbf{W})
	+ (1 - y_i) \log \textrm{P}(y_i = 0 | \mathbf{X};\mathbf{W})\right),
	\label{eq:loss}
\end{equation}
where $y_i$ is the ground truth label of a voxel (i.e. 1 for particle, 0 for background), $\textrm{P}(y_i | X;\mathbf{W})$ is the predicted probability of the voxel belonging to either a particle or the background, given input $\mathbf{X}$ and the network weights $\mathbf{W}$, and $N$ represents the total number of voxels in a given batch.
\\

We use the Adam optimizer to update the network parameters. 
For training the DSN, while the GTN is initialized with Xavier random weights in all cases, two different initialization strategies are employed for the experts within the DSN, as discussed in Section~\ref{sec:tr_init_strategy}.
In the first case, each expert within the DSN is initialized with the weight of a pretrained expert DNN on a specific particle density from $\rho = \{3.2, 6.41, 12.82\}\times 10^4$ particles/$\mu$L, respectively. 
The pre-training of these expert DNNs follows the same procedure for training the expert networks.
With this initialization, the DSN was trained with a learning rate of $10^{-5}$ with a batch size of 1 for about $30\tn{K}$ iterations.
Each iteration takes about 0.62 seconds, and the training time for 1 epoch is around 5 hours.
After training, we perform model selection using a validation set.
For the expert initialized DSN, the best model is trained for only one epoch.
In the second initialization strategy, no pre-training is performed, and all the experts within the DSN are initialized using Xavier random initialization.
In this case, the DSN is trained using the same learning parameters as the first initialization strategy. 
Due to the random initialization, the best model is trained for four epochs, which converges with about $120\tn{K}$ iterations and 20 hours.
Additionally, L2 weight decay was observed to improve the results from the DSN. 
Thus, L2 weight regularization was employed for the DSN training, with a regularization parameter $\gamma=1e^{-6}$, making the effective loss function for the DSN as $\mathcal{L(\mathbf{W})} + \gamma ||\mathbf{W}||^2_2$. 
\\

For our computations, we use an Intel Xeon E5-1630 v4 3.7GHz processor with 128 GB RAM and an Nvidia Quadro RTX-8000 GPU. 

\subsection{Expert and generalist network implementation}
\label{sec:Gnetwork}
As mentioned in Section~\ref{sec:G3x}, we implemented two generalist networks to compare with our DSN.
The first generalist network is the same as the expert network, as shown in Fig.~\ref{fig:s_Vnet}.
The difference is that when training this generalist, the data set contains different scattering data, as detailed in Section~\ref{sec:dataset}.
The parameters of each layer in this generalist are detailed in Table~\ref{fig:gen_parameters}.
We use this generalist as the first baseline because the synthesized DSN combines three different experts into a single V-net as shown in Fig.~\ref{fig:overview}.
After the synthesis shown in Fig.~\ref{fig:feature}(a), the DSN has the same number of learned features as this generalist (or expert), as shown in Fig.~\ref{fig:feature}(b).
By comparing this baseline generalist network, we aim to elucidate on how feature synthesis by the DSN improves the generalization capability.
\\

The second generalist network contains 3$\times$ number of network parameters, termed 3$\times$ generalist, as detailed in Section~\ref{sec:G3x}.
To implement the 3$\times$ generalist, the same V-net structure is maintained while additional numbers of channels are used, as detailed in Table~\ref{fig:3_gen_parameters}.
%
%
The number of channels used in different layers were heuristically optimized. 
The 3$\times$ generalist provides the second baseline for comparison with our DSN.
While they have the same number of trainable parameters in training, during inference, the DSN uses about 3$\times$ fewer features to achieve the high performance. 
This highlights that the dynamic feature synthesis enabled by the DSN structure makes the network more efficient to generalize as compared to increasing the network size in brute-force (i.e. in the 3$\times$ generalist). 
\\

Both generalist networks use Xavier random weight initialization.
For the baseline generalist, the network is trained for about 17K iterations with a learning rate of $10^{-5}$ and a batch size of 20. Each iteration takes about 2.90 seconds and the total training time for this network is 14 hours.
For the $3\times$ generalist, the network is trained for about 30K iterations (one epoch) with a learning rate of $10^{-4}$ and a batch size of 1.
The best model is selected based on the validation set.
Each iteration takes about 0.43 seconds and the total training time for this network is around 4 hours.
\\

The expert network also uses the Xavier random weight initialization and is trained for about 81K iterations with a learning rate of $10^{-4}$ and a batch size of 4.
Each iteration takes about 0.80 seconds, and the total training time for each network is about 18 hours.
\\

Both generalist networks and the expert networks are trained using the same loss function in Equation~\eqref{eq:loss}. L2 weight regularization is found to have no effect on their results.

\subsection{Experimental setup}
\label{sec:setup}
Our experimental setup for inline holography, shown in Fig.~\ref{fig:s_setup}, uses a linearly polarized HeNe laser ($\lambda=632.8$~nm, $500:1$ polarization ratio, Thorlabs HNL210L) that is collimated for illuminating the sample. 
A 4F system with a 20$\times$ objective lens (0.4 NA, CFI Plan Achro), and a $200$~mm tube lens is used to relay the field onto a CMOS sensor (FLIR GS3-U3-123S6M-C, pixel size $3.45~\mu$m, cover glass removed by Wilco imaging, Inc) for recording the holograms, each containing 1024$\times$1024 pixels.
The effective lateral pixel size is $172.5$~nm, which satisfies the Nyquist sampling requirement.
The sample consists of polystyrene microspheres suspended in deionized water, with nominal diameter $0.994~\mu \tn{m} \pm 0.021~\mu$m and refractive index $1.59$ (Thermofisher Scientific 4009A).
The sample is held in a quartz-cuvette with inner dimensions $40 \times 40 \times 0.5$~mm$^3$.
The front focal plane of the  objective lens was set slightly outside of the inner wall of the cuvette for hologram recording.
A shutter speed of $5~\tn{ms}$ was used and found to be sufficiently fast to avoid any motion artifacts from the moving particles.
The illumination beam diameter was kept less than the width of the cuvette, while larger than the CMOS sensor area to avoid edge artifacts.
With this setup, we acquire holograms for samples with approximate particle densities at $\rho = \{1.6, 3.2, 6.41, 12.82, 25.64\}\times 10^4$ particles/$\mu$L, corresponding to approximately 250, 500, 1000, 2000, 4000 particles in the imaged volume of size $176.64 \times 176.64 \times 500~\mu$m$^3$.
\\

\subsection{Multiple-scattering simulation} 
\label{sec:simulator}

Since the ground-truth particle locations for the experimental holograms are not known, we employ simulated data for the DSN training.
We simulate 3D samples with particle densities that approximately match the experimental data, including $\rho = \{1.6, 3.2, 6.41, 12.82\}\times 10^4$ particles/$\mu$L, containing 250, 500, 1000, 2000 particles in the simulation volume for our training set and different scattering conditions for our test set, more details about the data set are detailed in Section~\ref{sec:dataset}.
The particle locations are placed randomly using the Poisson disk random sampling~\cite{bridson2007}.
The particle diameter $D$ is 1.0~$\mu$m and the refractive index contrast $\Delta n$ between the particle and the background medium (water) is 0.26.
The size of each synthetic 3D sample is $176.64 \times 176.64 \times 500 ~\mu$m$^3$, corresponding to $1024 \times 1024 \times 4222$ voxels. 
The lateral field of view corresponds to the experimental hologram size.
The axial size matches with the internal depth of the cuvette used in the experiment.
The lateral step size is chosen to be $\delta x= \delta y = 172.5$~nm, which corresponds to the effective pixel size in the experiment.
The axial step size is chosen to be $\delta z=118.4$~nm, corresponding to $\lambda_{m}/4$ in order to accurately model multiple scattering~\cite{wang2021}, where $\lambda_{m}$ is the wavelength in the aqueous medium, i.e. $\lambda/1.33$, 1.33 being the refractive index of water.
For simulating the holograms given the synthetic particle volume, we use the BPM to accurately model the multiple scattering process since it demonstrates minimal discrepancy between the measured and simulated holograms and is highly computationally  efficient for simulating large-scale data with GPU acceleration~\cite{wang2021}.
The implementation detail of the BPM along with the accompanying open-source code is provided in \cite{wang2021}.
\\

\subsection{Holographic backpropagation}
\label{sec:holo-prop}
To generate the input to the DNN, we perform holographic backpropagation on each hologram to obtain an initial estimate of the 3D particle fields.
This 3D backpropagation corresponds to the minimum-norm solution for the 2D-to-3D reconstruction problem under the linear first Born approximation model~\cite{chen2015e}. 
The 3D holographic backpropagation $\mb{R}(x,y;z)$ is computed by numerically propagating the hologram $\mb{I}(x,y)$ ``backwards'' from the hologram plane to the object volume slice-by-slice, as follows:
\begin{equation}
    \mb{R}(x,y;z) = \mc{F}^{-1}
    \{
    \mc{F}
    \{
    \mb{I}(x,y)
    \}
    \cdot
    \mc{H}(u,v;z)
    \}
\end{equation}
where $\mc{F}\{\boldsymbol{\cdot}\}$ denotes the 2D Fourier transform and $\mc{F}^{-1}\{\boldsymbol{\cdot}\}$ its inverse, $x$ and $y$ the lateral spatial coordinates, $z$ the axial distance, and $u$ and $v$ the transverse spatial frequencies.
$\mc{H}(u,v;z)$ is the transfer function of the free-space Green's function $\mc{G}(x,y;z) = \exp(ikr)/r$, where $k=2\pi/\lambda_{m}$ is the wave number and $r = \sqrt{x^2+y^2+z^2}$, and $\mc{H}(u,v;z)$ is computed by taking the slice-wise 2D Fourier transform of the Green's function: $\mc{H}(u,v;z) = \mc{F}\{\mc{G}(x,y;z)\}$.
For the backpropagation, we set $\delta z = 5~\mu$m to approximately match the axial resolution of the experimental setup, determined by $\lambda_m / (1-\sqrt{1-\tn{NA}^2}) = 5.7~\mu$m.
This is because we do not attempt to localize particles with better accuracy than the system's axial resolution.
This produces a backpropagated volume with  $1024 \times 1024 \times 100$ voxels.
Since our problem is highly ill-posed due to the large dimensional mismatch between the object and measurement domains, the backpropagated volume contains significant scattering artifacts whose severity is directly proportional to the underlying particle density of the sample and increases at deeper depth slices, as depicted in Fig.~\ref{fig:quali} and Fig.~\ref{fig:s_bprop}. 
\\

\subsection{Data preprocessing}
\label{sec:preprocess} 
All DNN models in this work are trained using  supervised learning framework.
Therefore, in order to train a network to remove scattering artifacts from the 3D backpropagated volume, the corresponding ground truth object is also required.
The ground truth is difficult to obtain for our experimental data, instead we {\it only} use simulated data for the DNN training, for which the ground truth is available.
A remaining challenge is the large scale of the 3D imaging problem.
Each synthetic 3D sample contains $1024 \times 1024 \times 4222$ voxels, which is too large to be computed directly on our DNNs.
Simply dividing the depth into smaller sub-volumes would lose contextual information about the depth-dependent scattering artifacts clearly visible in the back-propagated volumes in Fig.~\ref{fig:quali} and Fig.~\ref{fig:s_bprop}. 
Instead, we down-sample the ground-truth volume with axial binning to form a smaller sized volume containing $1024 \times 1024 \times 100$ voxels.
For this purpose, we project each of the 4222 slices in the original synthetic 3D sample to the closest slice within the 100 slices in the ground-truth volume.
The resulting $1024 \times 1024 \times 100$ volume is used as the ground-truth label for the DNN training.
In addition, we pose the training as a detection problem with a binary ground truth, i.e. every voxel represents whether it belongs to a particle $\{1\}$ or the background~$\{0\}$.
Essentially, we do not attempt to reconstruct the actual refractive index of each particle but only aim to localize them as a compromise to the severe ill-posedness of the inverse problem.
\\

To preprocess the backpropagated volume before inputting it to the DNN, we first take the amplitude of each complex-valued backpropagated volume. 
Next, we normalized each volume by subtracting its mean and then dividing its standard deviation:  $\mb{\tilde{g}}=(\mb{g}-\mu_g)/{\sigma_g}$, where $\mb{\tilde{g}}$ is the preprocessed volume, $\mb{g}$ is the original volume, and $\mu_g$ and $\sigma_g$ are the mean and standard deviation of the volume, respectively.
\\

The input to the GTN is a preprocessed hologram. 
We normalize the hologram by $\mb{\tilde{I}} = ({\mb{I}-\mu_{h}})/{\sigma_{h}}$, where $\mb{\tilde{I}}$ is the preprocessed hologram, $\mb{I}_{h}$ is the original simulated or captured hologram, and $\mu_{h}$ and $\sigma_{h}$ are the mean and standard deviation of the hologram, respectively.
Two example holograms captured from different scattering densities and the intensity histograms before and after the normalization are shown in Fig.~\ref{fig:preprocessing}.
It is evident that, by this hologram normalization scheme, the GTN is forced to extract non-trivial differences in holograms captured at different scattering conditions in order to output distinct synthesis weights.
\\

\subsection{Simulated training and testing data sets}
\label{sec:dataset} 
To train each expert DNN for a specific condition (the refractive index is 0.26, the particle diameter is 1.0~$\mu$m, and the particle density is varied for different experts), we randomly generate 48 training and 2 validation pairs of the input backpropagation and the corresponding ground-truth volumes for each expert, each of which is $1024 \times 1024 \times 100$ voxels in size.
We further divide each volume to have a smaller lateral size while keeping the axial dimension fixed to make the end-to-end 3D computation feasible.
Specifically, the input to the DNN comprises $128 \times 128 \times 100$-voxel sub-volumes cropped from the full backpropagation volumes. 
The sub-volumes are cropped to have a $64 \times 64$-voxel lateral overlap among the neighboring patches for training, while no overlap for validation and testing.
In total, for each expert DNN, we generated 10800 sub-volumes for training, and 128 for validation.
\\

To train the generalist and the DSN, we generate 70 training and 4 validation pairs of $1024 \times 1024 \times 100$-voxel backpropagation and ground-truth volumes, corresponding to 15750 training, and 256 validation sub-volumes after performing the same volume division procedure.  
When training the DSN using each sub-volume, the matching preprocessed hologram patch ($128\times128$ pixels) is also fed into the GTN for predicting the synthesis weights $\alpha$.
The same data set is used for training the generalist and the DSN for fair comparison of their performance. 
The particle refractive index contrast is 0.26 and the diameter is $1.0~\mu$m, which is the same as the expert network and the particles used in our experiment.
The particle densities ($\rho\times 10^4$ particles/$\mu$L) and the number of volumes ($n$) at each density used for training the generalist and the DSN include:
$\rho=1.6$ ($n$ = 10), $\rho=3.2$ ($n$ = 20), $\rho=6.41$ ($n$ = 20), and $\rho=12.82$ ($n$ = 20).
The validation data set includes $\rho=1.6$ ($n$ = 1), $\rho=3.2$ ($n$ = 1), $\rho=6.41$ ($n$ = 1), $\rho=12.82$  ($n$ = 1).
We empirically optimized the choice of training data.
We chose to not include higher densities like $\rho = 25.64$ for the training, since we observe that the data does not sufficiently benefit the DSN or generalist network training due to the poor effective ``signal-to-noise'' in the backpropagated volumes. 
%
\\

First, we test all our trained DNNs on a variety of ``seen" scattering conditions.
We generate 40 testing pairs of backpropagated and ground-truth volumes with $1024 \times 1024 \times 100$ voxels.
The refractive index contrast is 0.26, the particle diameter is 1.0~$\mu$m, and the four different particle densities are the same as the training set. 
The particle densities ($\rho\times 10^4$ particles/$\mu$L) used for this testing task and the number of testing pairs ($n$) include:
$\rho=1.6$ ($n$ = 10), $\rho=3.2$ ($n$ = 10), $\rho=6.41$ ($n=10$), $\rho=12.82$ ($n=10$).
\\

To further test the generalization capability of the DSN and other DNNs, we also test our them on several ``unseen'' scattering conditions, including different particle densities, sizes and refractive index contrasts. The testing results are summarized in Fig.~\ref{fig:unseen}.
These testing cases for generalization are categorized into the following eight groups:

(1) Unseen particle densities ($\rho\times 10^4$ particles/$\mu$L) with seen refractive index contrast (0.26) and particle diameter (1.0~$\mu$m) and the number of testing pairs ($n$) include:
$\rho=2.24$ ($n$ = 10), $\rho=4.81$ ($n$ = 10), $\rho=9.61$ ($n=10$), $\rho=19.23$ ($n=10$) and $\rho=25.64$ ($n=10$).

(2) Unseen particle diameter ($D~\mu$m) with seen refractive index contrast (0.26) and density ($\rho=6.41$) include:
$D=0.30$ ($n$ = 10), $D=0.60$ ($n$ = 10), $D=1.40$ ($n=10$), $D=1.80$($n=10$).

(3) Unseen particle refractive index contrast ($\Delta n$) with seen particle diameter ($1.0~\mu$m) and density ($\rho=6.41$) include:
$\Delta n=0.01$ ($n$ = 10), $\Delta n=0.05$ ($n$ = 10), $\Delta n=0.10$ ($n$ = 10),  $\Delta n=0.20$ ($n=10$), $\Delta n=0.30$, $\Delta n=0.40$ and $\Delta n=0.50$.

(4) Unseen particle size and Unseen refractive index contrast ($D - \Delta n$) with ``seen" particle density ($\rho=6.41$) include:
$D=0.30-\Delta n=0.10$ ($n=10$), $D=0.30-\Delta n=0.50$ ($n=10$), $D=0.60-\Delta n=0.10$ ($n=10$), $D=0.60-\Delta n=0.50$ ($n=10$), $D=1.30-\Delta n=0.10$ ($n=10$), $D=1.30-\Delta n=0.50$ ($n=10$), $D=1.50-\Delta n=0.10$ ($n=10$), $D=1.50-\Delta n=0.50$ ($n=10$).

(5) Unseen particle refractive index contrast ($\Delta n=0.10$), unseen particle size (0.60~$\mu$m) and unseen particle densities ($\rho\times 10^4$ particles/$\mu$L) include:
$\rho=2.24$ ($n$ = 10), $\rho=4.81$ ($n$ = 10), $\rho=9.61$ ($n$ = 10), $\rho=19.23$ ($n=10$) and $\rho=25.64$ ($n$ = 10).

(6) Volumes containing particles with random refractive index contrasts that follows a uniform distribution $U(\Delta n_c,\delta_n)$ with a fixed central refractive index contrast ($\Delta n_c = 0.20$) and different amounts of variations $\delta_n$, and at a seen particle density ($\rho=6.41$) and particle size ($D = 1.0~\mu$m) include: 
$\delta_n=\pm5.0\%$ ($n$ = 10), $\delta_n=\pm12.5\%$ ($n$ = 10), $\delta_n=\pm25.0\%$ ($n$ = 10) and $\delta_n=\pm50.0\%$ ($n$ = 10), where $\delta_n$ is measured by the ratio ($\%$) between the index variation range and the central refractive index contrast.

(7) Volumes containing particles with random diameters that follows a uniform distribution $U(D,\delta_D)$ and different amounts of variations $\delta_D$, and with a fixed central diameter ($D = 1.0~\mu$m), an unseen  refractive index contrast ($\Delta n = 0.20$) and a seen particle density ($\rho=6.41$) include:
$\delta_D=\pm9.0\%$ ($n$ = 10), $\delta_D=\pm17.5\%$ ($n$ = 10), $\delta_D=\pm26.0\%$ ($n$ = 10) and $\delta_D=\pm34.5\%$ ($n$ = 10), where $\delta_D$ is measured by the ratio ($\%$) between the diameter variation range and the central diameter.

(8) Volumes containing particles with random refractive index contrasts and random particle diameters, and with a seen density ($\rho=6.41$).
The refractive index contrast follows $U(0.20,\delta_n)$.
The particle diameter follows $U(1.0~\mu\mathrm{m},\delta_D)$.
This testing set, labeled as $\delta_n-\delta_D$, include: 
$\delta_n=\pm5.0\%-\delta_D=\pm9.0\%$ ($n$ = 10), $\delta_n=\pm12.5\%-\delta_D=\pm17.5\%$ ($n$ = 10), $\delta_n=\pm25.0\%-\delta_D=\pm26.0\%$ ($n$ = 10), $\delta_n=\pm50.0\%-\delta_D=\pm34.5\%$ ($n$ = 10).
%

\subsection{Performance evaluation metrics} 
\label{sec:metrics}

The performance of the DNN on the simulated data is quantified using the Jaccard index (JI). 
To do so, we first discuss the procedure of assigning True Positive (TP), False Positive (FP), and False Negative (FN) labels to the reconstructed particles, and then describe the computation of the JI.
TP represents a correctly localized particle, FP represents a falsely detected particle in the reconstruction that is not present in the ground truth, and FN represents a missed particle which is not detected in the reconstruction but is present in the ground truth.
The computation involves segmentation of the network output, followed by centroid detection, and the comparison with the ground-truth particle locations, as detailed below.
\\

The output from the DNN is a 3D probability map representing the likelihood of each voxel belonging to either a particle or the background.
To process the network's output, first we perform slice-wise automatic thresholding using Otsu's method~\cite{otsu1979} to obtain a 3D binary map.
All clusters composed of less than 10 voxels are discarded as noise, similar to our previous work~\cite{tahir2019}, since they are significantly smaller than the expected particle size of 21 voxels.
Second, we perform centroid detection for all the reconstructed particles.
Third, we perform pairwise matching between the reconstructed  and the ground-truth centroids by computing  the pairwise distances and solving a linear assignment problem~\cite{burkard1999}. 
Fourth, we assign each reconstructed particle a label from TP, FP, and FN based on the following criterion. 
TP is assigned to a particle if it is within an elliptical proximity volume of the matching ground-truth particle.
We choose the axial dimension of the proximity volume to be $12~\mu$m, which is roughly $2\times$ the axial resolution ($5.7~\mu$m) of our system.
Even though the lateral resolution of our  system is $\lambda_m / (\tn{NA}) = 1.2~\mu$m, empirically the localization performance decreases as depth increases. %
To account for this, we heuristically choose the lateral dimensions of the proximity volume to be $4 \times$ the training particle size of $1\mu$m.
Thus, the proximity volume is an ellipse of dimensions $4 \times 4 \times 12~\mu$m$^3$.
FP is assigned to a particle if it either does not get matched with any ground-truth particle in the third step, or it is outside the proximity volume of the matching ground-truth particle.
FN is assigned to a particle if it is in the ground truth volume but is not matched with any reconstructed particle in the third step, or the matched reconstructed particle is outside the proximity volume.
Finally, using these assigned labels, the JI is computed as 
\begin{equation}
    \tn{JI} = \frac{\tn{TP}}{\tn{TP}+\tn{FP}+\tn{FN}},
\end{equation}
which measures the similarity between the reconstructed and ground truth particle locations. 
The JI is computed for groups of 10 axial slices (i.e. every $50 \mu$m) in the reconstructed 3D volumes.
\\


\subsection{UMAP visualization} 
\label{sec:UMAP}
Uniform Manifold Approximation and Projection (UMAP) is the state-of-the-art unsupervised dimensionality reduction technique that models the entire data set into a low dimensional manifold by learning the underlying topological structure contained in the original high-dimensional data~\cite{mcinnes2018}.
For UMAP visualization, we consider each data (e.g. a hologram) as a single vector, which is mapped to a single point in a 2D manifold learned by the UMAP algorithm.
Applying UMAP to a whole data set can provide insights into the statistical distribution of the high-dimensional data set by visualizing the learned low-dimensional manifold.
In the UMAP manifold, statistical similar samples are clustered in close proximity, while dissimilar samples are separated. 
Its utility for visualizing underlying correlations in scattering measurements have been demonstrated in our recent work~\cite{li2021}.
\\

For the UMAP visualization in Fig.~\ref{fig:umap}(a), we use 5120 non-overlapping preprocessed hologram patches of $128 \times 128$ pixels from 80 preprocessed holograms, distributed equally among the particle densities of $\rho = \{1.6, 3.2, 6.41, 12.82\}\times 10^4$ particles/$\mu$L. 
From this figure, we can see that even after normalization, the holograms from different particle densities are still clustered based on the underlying density.
%

\subsection{Statistical analysis of synthesis weights} 
\label{sec:stats_synweight}
We plot the bar chart of output weights given by GTN for different scattering conditions.
In Fig.~\ref{fig:umap}(b), for particles with refractive index contrast 0.26 and  diameter $1.0~\mu$m, we used in-total 1728 non-overlapping hologram patches ($128 \times 128$ pixels) cropped from 27 preprocessed holograms.
For each particle density, three preprocessed holograms (corresponding to 192 patches) are used. 
The following particle densities are used $\rho = \{1.6, 3.34, 3.2, 4.81, 6.41, 9.61, 12.82, 19.23, 25.64\}\times~10^4$ particles/$\mu$L.
In Fig.~\ref{fig:umap}(c), for particles with density $\rho=6.41\times10^4$ particles/$\mu$L and diameter $1.0~\mu$m, we used in-total 1536 non-overlapping hologram patches ($128 \times 128$ pixels) cropped from 24 preprocessed holograms.
For each refractive index contrast, three preprocessed holograms are used.
The following refractive index contrasts are used $\Delta n = {0.01, 0.05, 0.10}$, ${0.20, 0.26, 0.30, 0.40, 0.50}$.
In Fig.~\ref{fig:umap}(d), for particles with density $\rho=6.41\times~10^4$ particles/$\mu$L and refractive index contrast 0.26, we used in-total 960 non-overlapping hologram patches ($128 \times 128$ pixels) cropped from 15 preprocessed holograms.
For each particle size, three preprocessed holograms are used.
The following particle sizes are used $\Delta D = {0.30, 0.60, 1.00, 1.30, 1.50}$.
In Fig.~\ref{fig:umap}(e), for particles with refractive index contrast 0.20 and particle diameter $0.6~\mu$m, we use in-total 1728 non-overlapping hologram patches ($128 \times 128$ pixels) cropped from 27 preprocessed holograms.
For each density, three preprocessed holograms (corresponding to 192 patches) are used.
The following densities are used from $\rho = \{1.6, 3.34, 3.2, 4.81, 6.41, 9.61, 12.82, 19.23, 25.64\}\times10^4$ particles/$\mu$L.
\\

\section*{Acknowledgements} 
The authors acknowledge funding from National Science Foundation (1813848, 1846784).
The authors acknowledge Boston University Shared Computing Cluster for proving the computational resources.

\section*{Author contributions}
W.T and L.T. conceived the idea. W.T. prototyped the method and conducted initial experiments. H.W. refined the method and conducted additional experiments. All authors participated in the writing of the paper.

\section*{Conflict of interest}
The authors declare no competing interests.

\section*{Data availability}
The neural network and the data set used in this work are available at \url{https://github.com/bu-cisl/DynamicSyntesisNetwork}.

\newpage


\clearpage
\newpage

\renewcommand{\thefigure}{S\arabic{figure}}
\setcounter{figure}{0}

\pagenumbering{arabic} 

\centering
\section*{{\normalfont Supplementary information for:}\\Adaptive 3D descattering with a dynamic synthesis  network}

Waleed Tahir$^{1,\star}$, {Hao Wang}$^{1,\star}$, Lei Tian$^{1,2,*}$
\\

[1] Department of Electrical and Computer Engineering, Boston University, Boston, MA 02215, USA.
\\

[2] Department of Biomedical Engineering, Boston University, Boston, MA 02215, USA.
\\
$\star$ Equal contribution.
* Correspondence: leitian@bu.edu, Tel.: (617) 353-1334

\newpage

\begin{figure}
	\centering
	\includegraphics[width=1\linewidth]{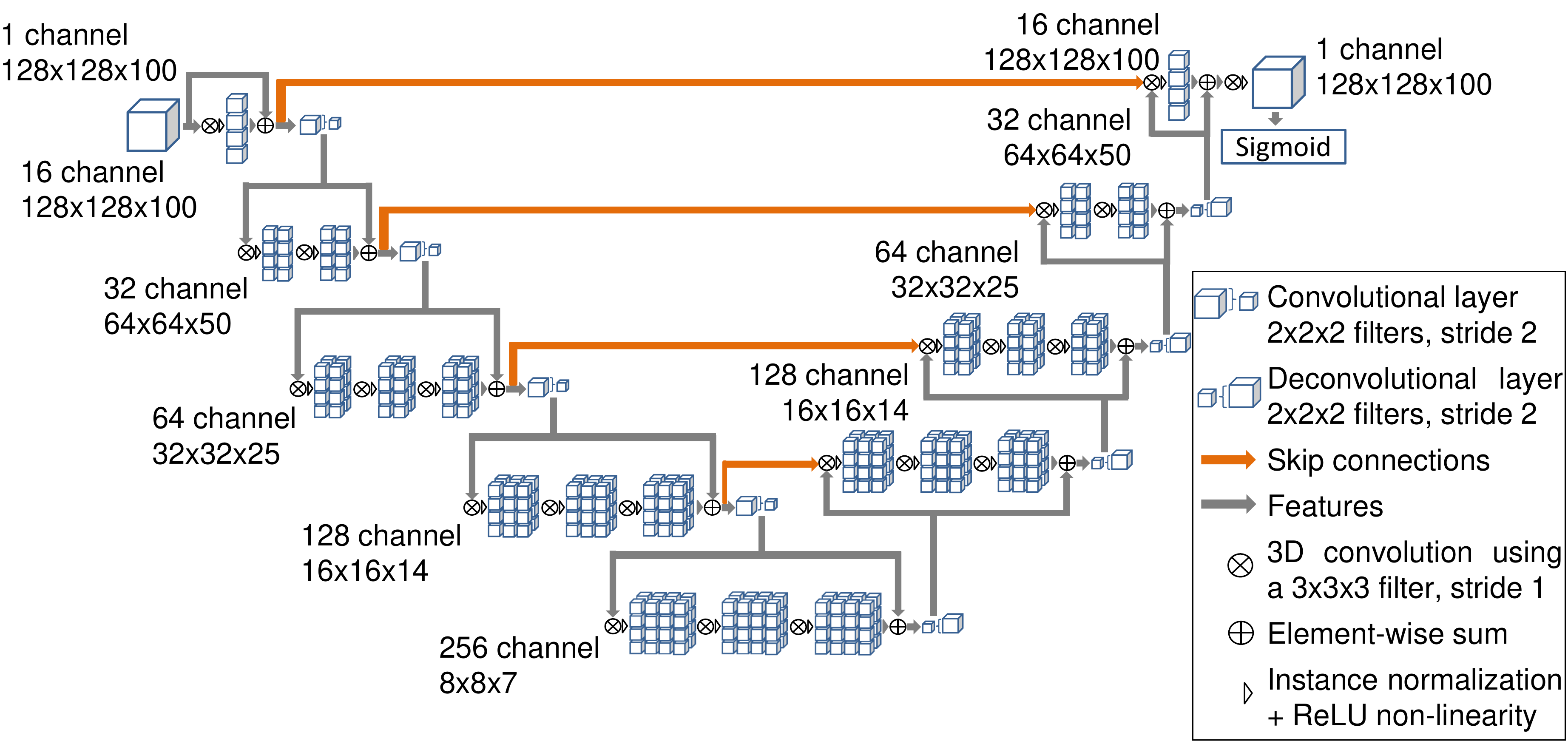}
	\caption{\textbf{Network architecture of the expert DNN.} The expert DNN builds on the V-net framework, which is a 3D DNN with an encoder-decoder framework. The encoder extracts multi-scale features using 3D convolution kernels of size $3\times3\times3$ with stride 1. Each layer in the encoder decreases the spatial dimensions by $2\times$ while doubling the number of channels. Spatial feature maps at different scales from the encoder are forwarded to the decoder via skip connections for preserving high resolution information. The bottleneck layer contains the ``latent code'' of size $8\times8\times7$ with 256 channels. The decoder also contains multiple 3D convolution layers with $3\times3\times3$ convolution kernels, and convolutional upsampling, with the additional incorporation of high resolution features from the skip connections. The single-channel output from the decoder is converted to a probability map using the sigmoid layer, in which each voxel value represents the likelihood of the voxel belonging to a particle (vs. belonging to the background).}
	\label{fig:s_Vnet}
\end{figure}

\begin{figure}
	\centering
	\includegraphics[width=0.75\linewidth]{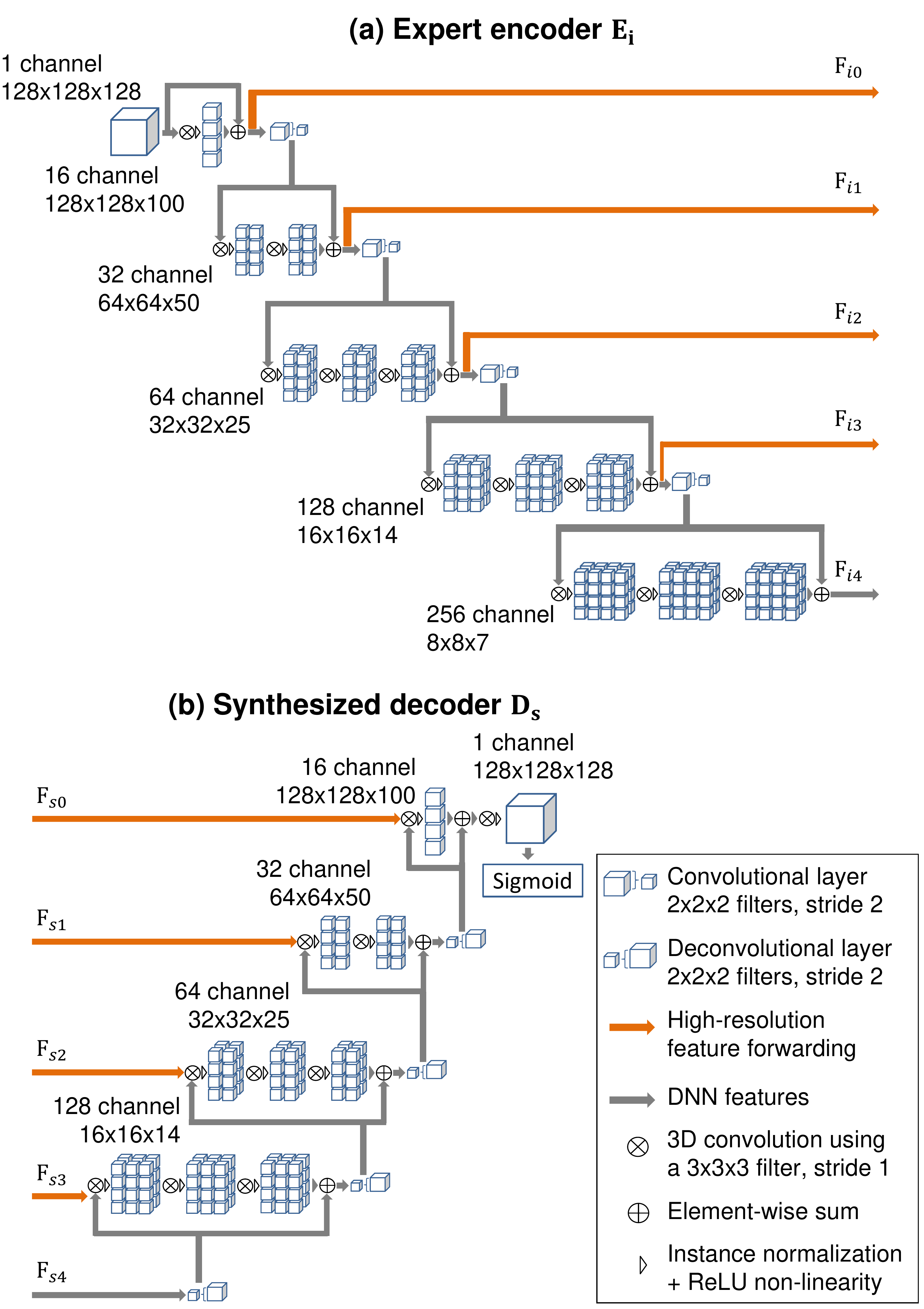}
	\caption{\textbf{Network architecture of the expert encoder $\tn{E}_i$ and synthesized decoder $\tn{D}_s$ within the DSN.} (a) The architectures of all the encoders $\tn{E}_i$ ($i \in \{1,2,3\}$) are identical, and are derived from a modified V-net. 
	The encoders extract features from the 3D input, which are then used by the decoder for generating the DNN output. Since V-net uses multi-scale feature forwarding by skip connections, we record the encoder extracted features at each spatial scale (denoted by $\tb{F}_{i0}$--$\tb{F}_{i3}$), together with the latent feature map $\tb{F}_{i4}$. This combined set of feature maps is labeled as $\tb{F}_{i} = [\tb{F}_{i0};\tb{F}_{i1};\tb{F}_{i2};\tb{F}_{i3};\tb{F}_{i4}]$ associated with the corresponding encoder $\tn{E}_{i}$. (b) The architecture of the synthesized decoder $\tn{D}_s$ is derived from the modified V-net decoder. Each network parameter in $\tn{D}_s$ is a weighted sum of the corresponding parameters in the three expert decoders $\tn{D}_i$: $\tn{D}_{s} = \sum_{i=1}^{3}\mathit{\alpha_i}\tn{D}_{i}$. Once $\tn{D}_s$ is synthesized, it decodes the corresponding synthesized multi-scale feature maps $\tb{F}_s = [\tb{F}_{s0};\tb{F}_{s1}\tb{F}_{s2};\tb{F}_{s3};\tb{F}_{s4}]$ for generating the output, where $\tb{F}_{sx} = \sum_{i=1}^{3} \alpha_i\tb{F}_{ix}$, and $x \in \{0,1,2,3,4\}$ indices the feature set.
	        }
	\label{fig:s_exp_struc}
\end{figure}

\begin{figure}
	\centering
	\includegraphics[width=1\linewidth]{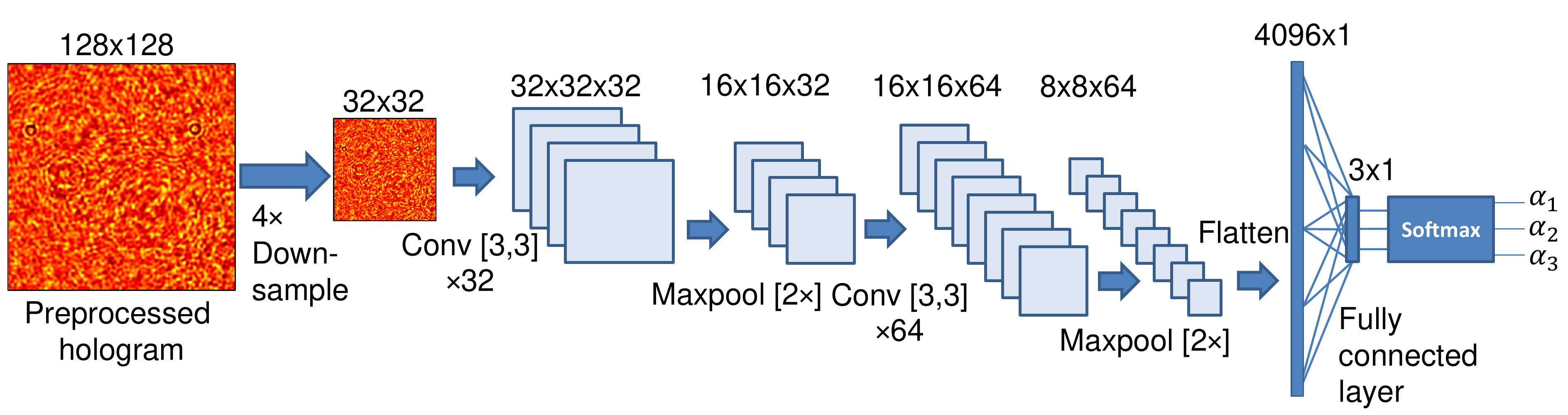}
	\caption{\textbf{Gating network structure.} 
	The GTN follows the VGG structure to predict the synthesis weights $\alpha_{i}$. 
	The input hologram is first downsampled by 4$\times$ by selecting every 4th pixel along each axes.
	Next, the GTN extracts multi-scale spatial features from the 2D hologram using two 2D convolutional layers, each containing a convolution with a bank of $3\times3$ kernels, followed by $2\times$ maxpooling to decrease the spatial dimensions. The first layer contains 32 channels; the second layer contains 64 channels.
	After the convolutional layers, the $8\times8\times64$ feature map is flattened and then passed through a fully connected layer.
    The output is a $3\times1$ vector representing the three synthesis weights $\{\alpha_{1}, \alpha_{2}, \alpha_{3}\}$. 
    The sum of the weights are enforced to be unity, i.e. $\sum_{i=1}^3 \alpha_{i} = 1$ through the use of the softmax nonlinear activation function at the last layer.
	        }
	\label{fig:s_GTN}
\end{figure}
\clearpage

\begin{figure}
	\centering
	\includegraphics[width=1\linewidth]{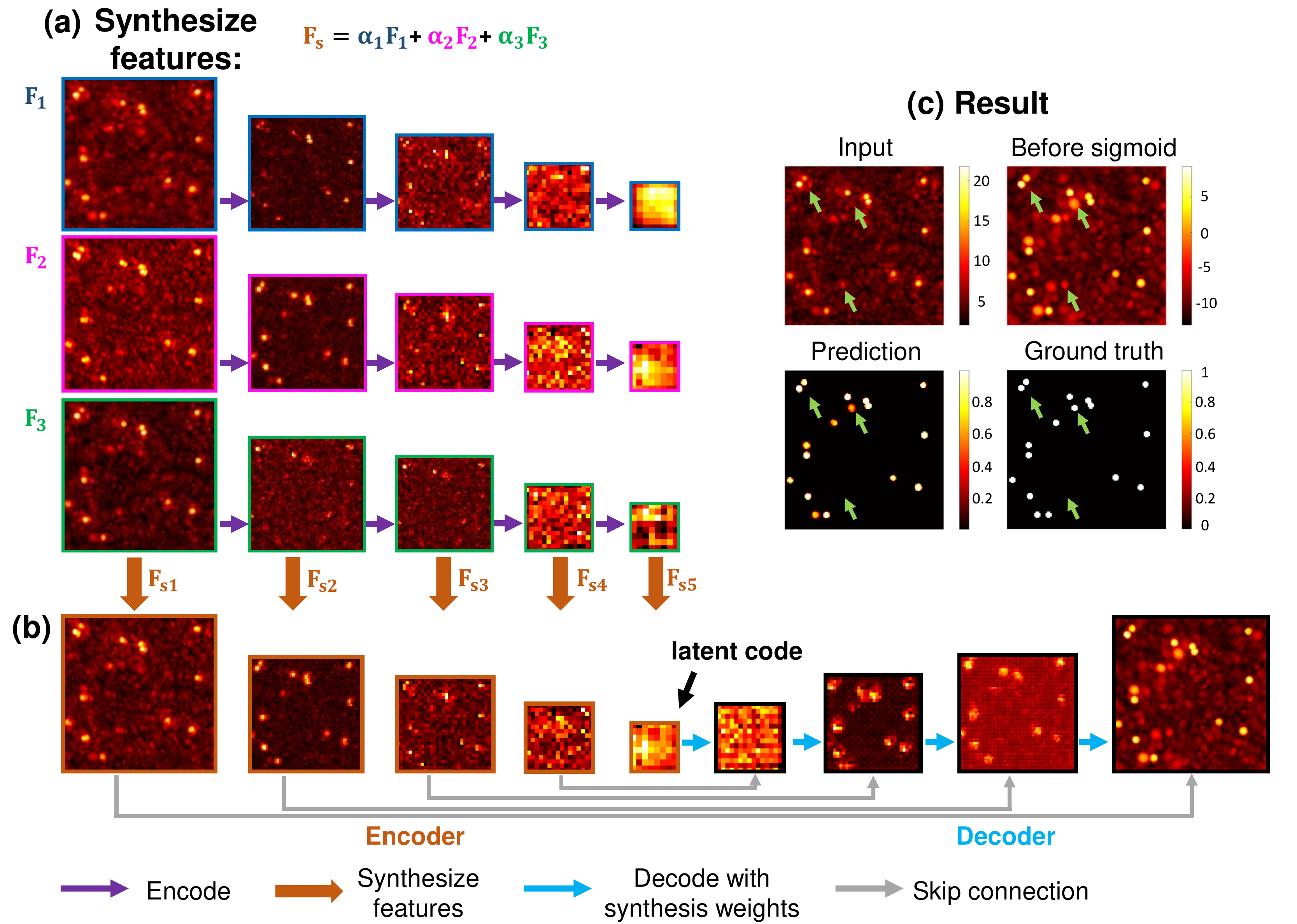}
	\caption{\textbf{Schematic diagram of the feature map flow in the DSN.} 
	We show the maximum $z$-projection of an example 3D feature map. 
	(a) The feature maps extracted by the three expert encoders. 
	(b) The feature maps in the synthesized DSN. 
	The synthesized encoder features are directly concatenated to the decoder by the skip connections.
	(c) The input and output of the DSN. 
	}
	\label{fig:feature}
\end{figure}
\clearpage

\begin{figure}
	\centering
	\includegraphics[width=1\linewidth]{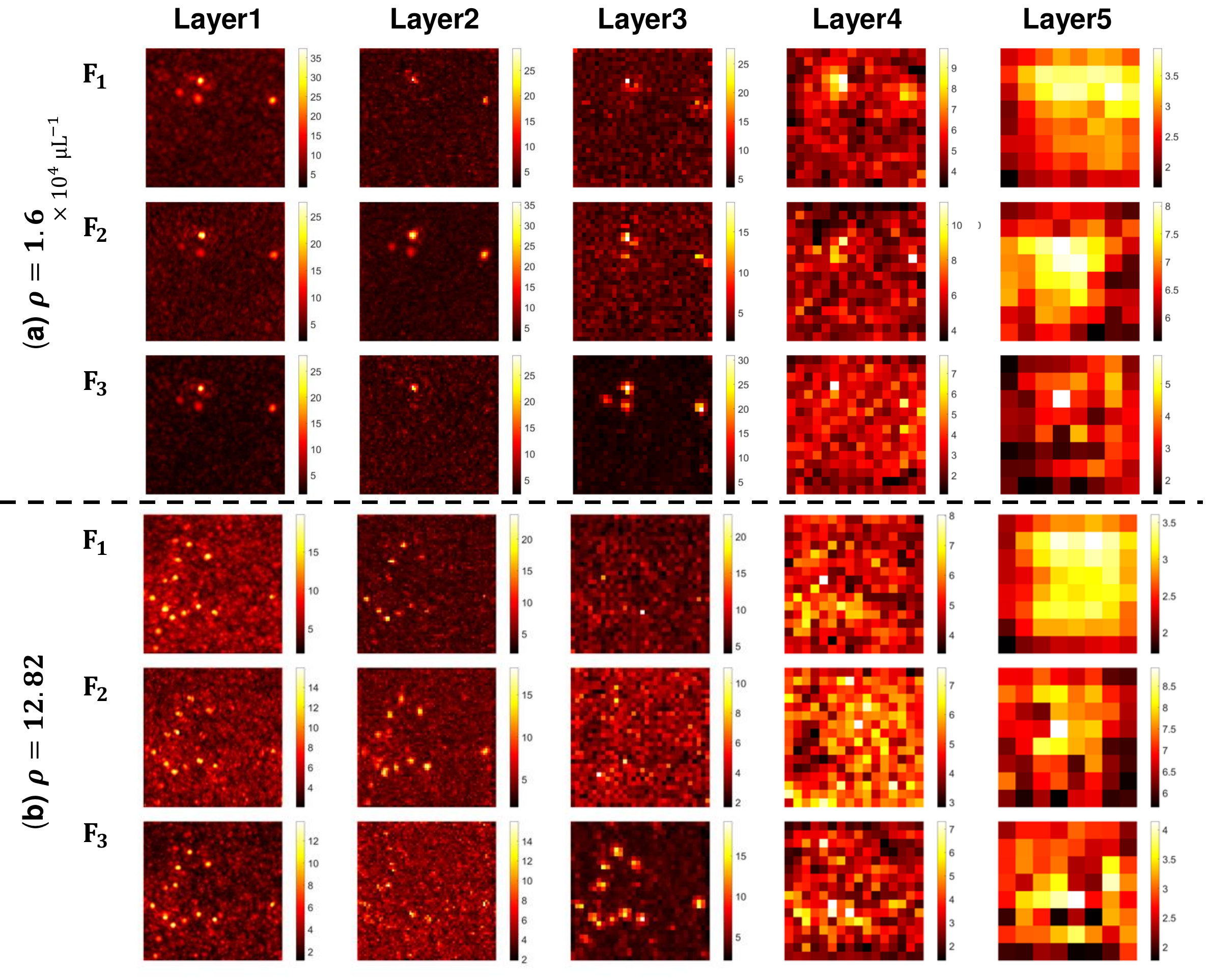}
	\caption{\textbf{DSN Feature maps at different particle densities.}
	The feature maps (shown in $z$-projects) of a low density and a high density are shown. 
	$\tb{F}_{1}, \tb{F}_{2}, \tb{F}_{3}$ represent the three expert encoders within the DSN. 
	Layer1--Layer5 represent the encoder layers within the DSN, the same as Fig.~\ref{fig:feature}(a).
}
	\label{fig:feature_diff_scatter}
\end{figure}
\clearpage

\begin{figure}
	\centering
	\includegraphics[width=1\linewidth]{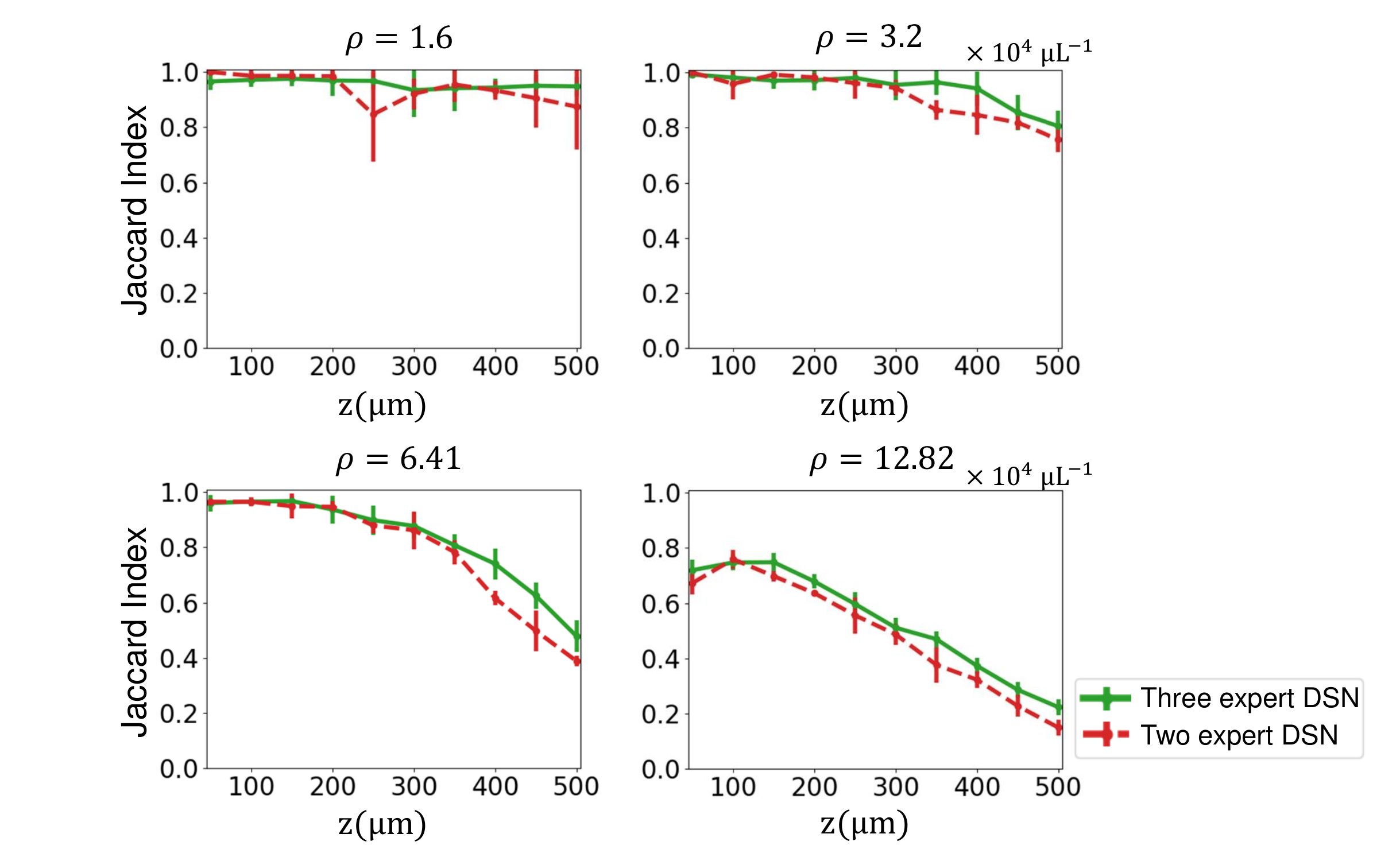}
	\caption{\textbf{Performance comparison: two vs three expert encoder-decoder pairs within the DSN.} 
		We compare the performance of the DSN with two (in dashed red) and three (in solid green) experts on the test set which has the same particle size, refractive index contrast and density as the training set. It is evident that the three-expert DSN performs better especially for higher particle densities ($\rho \geq 3.2 \times 10^4$ particles $\upmu \mr{L}^{-1}$).
		The results highlight the significance of the extra degrees of freedom provided by the additional expert within the DSN.}
	\label{fig:s_2v3}
\end{figure}

\clearpage

\begin{figure}
	\centering
	\includegraphics[width=0.85\linewidth]{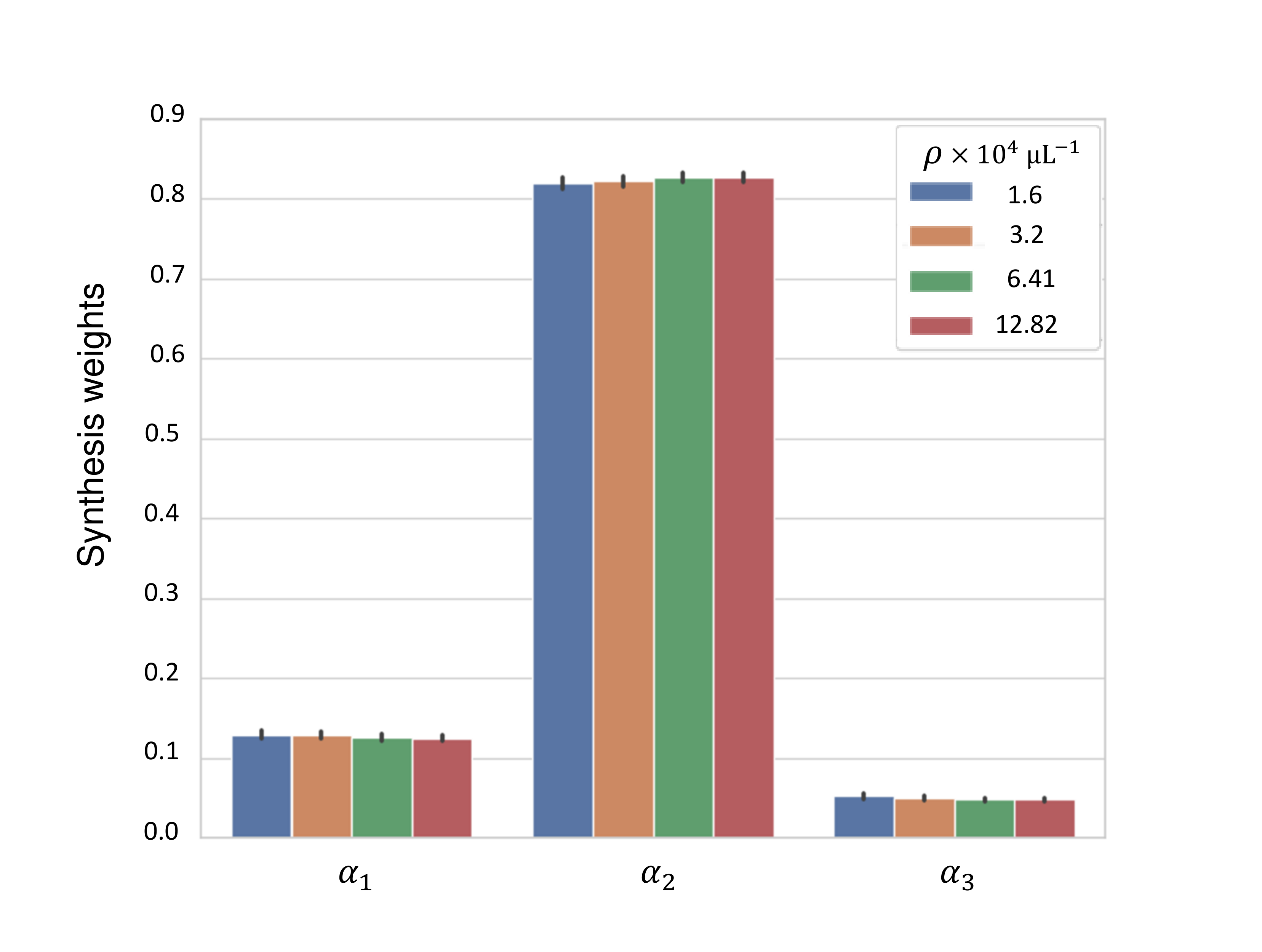}
	\caption{\textbf{The synthesis weights with Xavier-random initialization of the experts.}
	The synthesis weights of four different scattering densities ($\rho$) predicted by the GTN, co-trained with the random initialized experts, are shown for the simulated testing set (refractive index is 0.26, particle diameter is $1.0~\upmu $m).
	Different expert DNNs have different weights $\alpha$, which suggest that they contribute differently to the descattering.
	The larger values of $\alpha_{2}$ indicate the major contributions of $\tb{F}_{2}$ and $\tn{D}_{2}$ to the DSN. 
	The small values of $\alpha_{3}$ indicates the fine-tuned contributions from $\tb{F}_{3}$ and $\tn{D}_{3}$ to the DSN.
	The synthesis weights are consistent for a given density and tailored to each input, as quantified by the mean and standard deviation for each case.
	The general trend of the synthesis weights are similar to those obtained from the DSN using the pre-trained weight initialization scheme, shown in the Figure~\ref{fig:umap}.
	As the particle density increases, $\alpha_{1}, \alpha_{3}$ decreases while $\alpha_{2}$ increases, which indicates that $\tn{E}_{2}$ are important for the DSN to adapt to higher density cases. 
	The differences in the weight values compared with Figure~\ref{fig:umap} are expected from the different initialization schemes, the stochastic training process, and the severe ill-posedness of problem.
	}
	\label{fig:s_xavier_act}
\end{figure}
\clearpage

\begin{figure}
	\centering
	\includegraphics[width=1\linewidth]{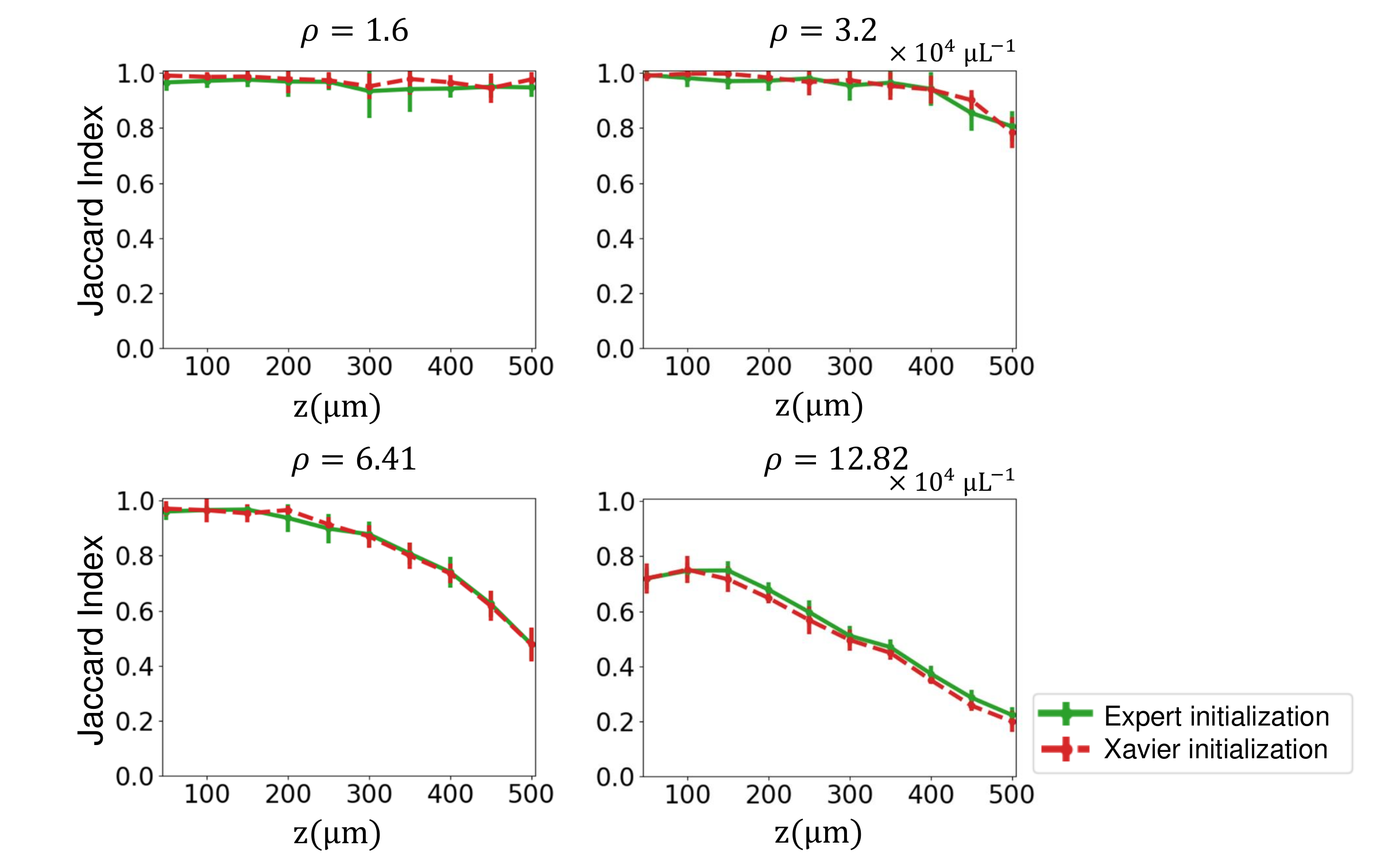}
	\caption{\textbf{Performance comparison: expert vs Xavier-random initialization of the encoders and decoders within the DSN.}
	Particle localization performance is quantitatively compared between the DSNs that are trained using two initialization schemes, including the pretrained expert weights (solid green) and Xavier random weights (dashed red) using the Jaccard Index (JI). 
	Each subplot indicates the results on the test set (refractive index is 0.26, particle diameter is $1.0~\upmu$m) at the particle density labeled above each plot.  
	The two initialization schemes provide almost the same performance.
	}
	\label{fig:s_rvi}
\end{figure}

\clearpage

\begin{figure}
	\centering
	\includegraphics[width=0.8\linewidth]{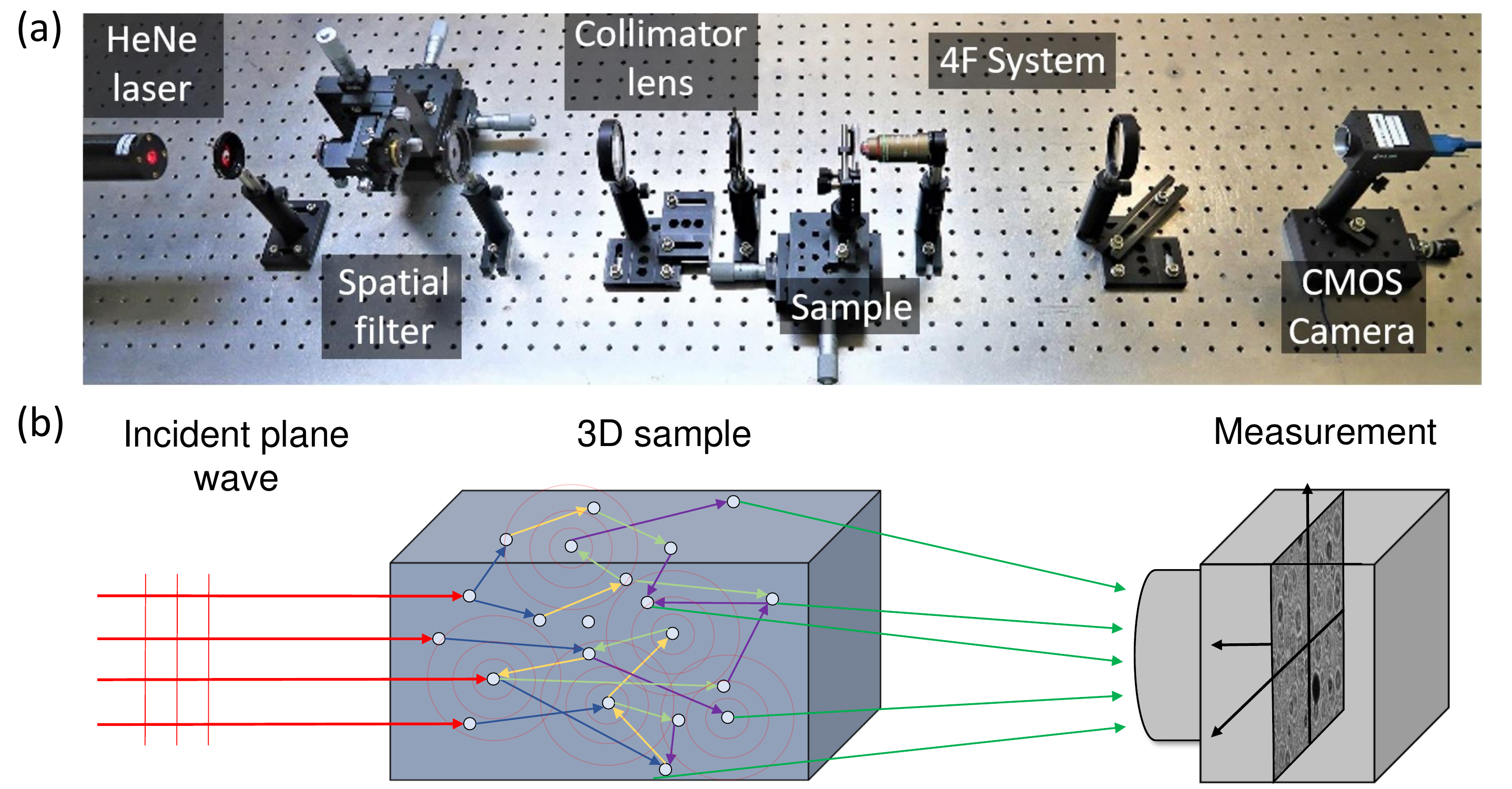}
	\caption{\textbf{Experimental setup and schematics.}
	(a) An inline holography setup consists of a collimated HeNe laser (632.8 nm, 500:1 polarization ratio, Thorlabs HNL210L) for illumination and a 4F system consisting of a 20$\times$ objective lens (0.4 NA, CFI Plan Achro) and a 200 mm tube lens for imaging. 
	A CMOS sensor (FLIR GS3-U3-123S6M-C) is used to record the holograms.  
	The 3D sample consists of polystyrene microspheres with diameter 0.994$\pm$0.021~$\upmu$m (Thermofisher Scientific 4009A) suspended in water held in a quartz-cuvette with inner dimensions 40 mm $\times$ 40 mm $\times$ 0.5 mm.
	(b) A plane-wave is incident on the 3D sample containing distributed particles. 
	The field undergoes multiple scattering and then propagates to the hologram plane. 
		}
	\label{fig:s_setup}
\end{figure}

\begin{figure}
	\centering
	\includegraphics[width=1\linewidth]{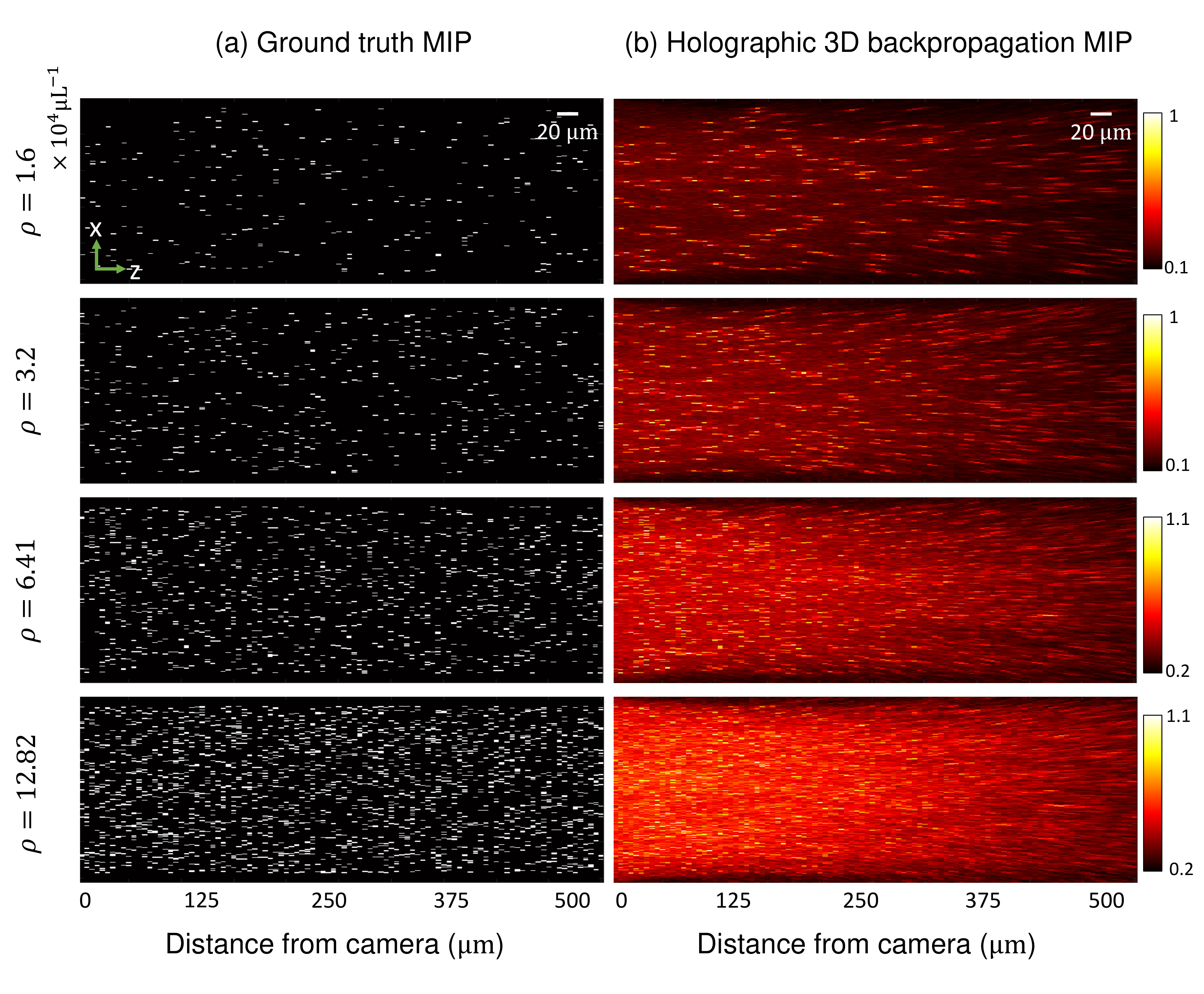}
	\caption{\textbf{Scatterer-density and depth-dependent artifacts in representative holographic 3D backpropagation volumes used in the DSN training.} 
	Maximum intensity $y-$projections of (a) ground-truth volume (particles shown in white, background in black), and (b) holographically backpropagated volume. 
	Characteristic scatterer-density and depth-dependent artifacts are clearly visible.  
	As the particle density increases, more severe scattering artifacts throughout the volume are shown.
	More elongation and reduced intensity in the backpropagated particle traces are observed at deeper depths due to reduced effective light collection angular range.}
	\label{fig:s_bprop}
\end{figure}

\begin{figure}
	\centering
	\includegraphics[width=1\linewidth]{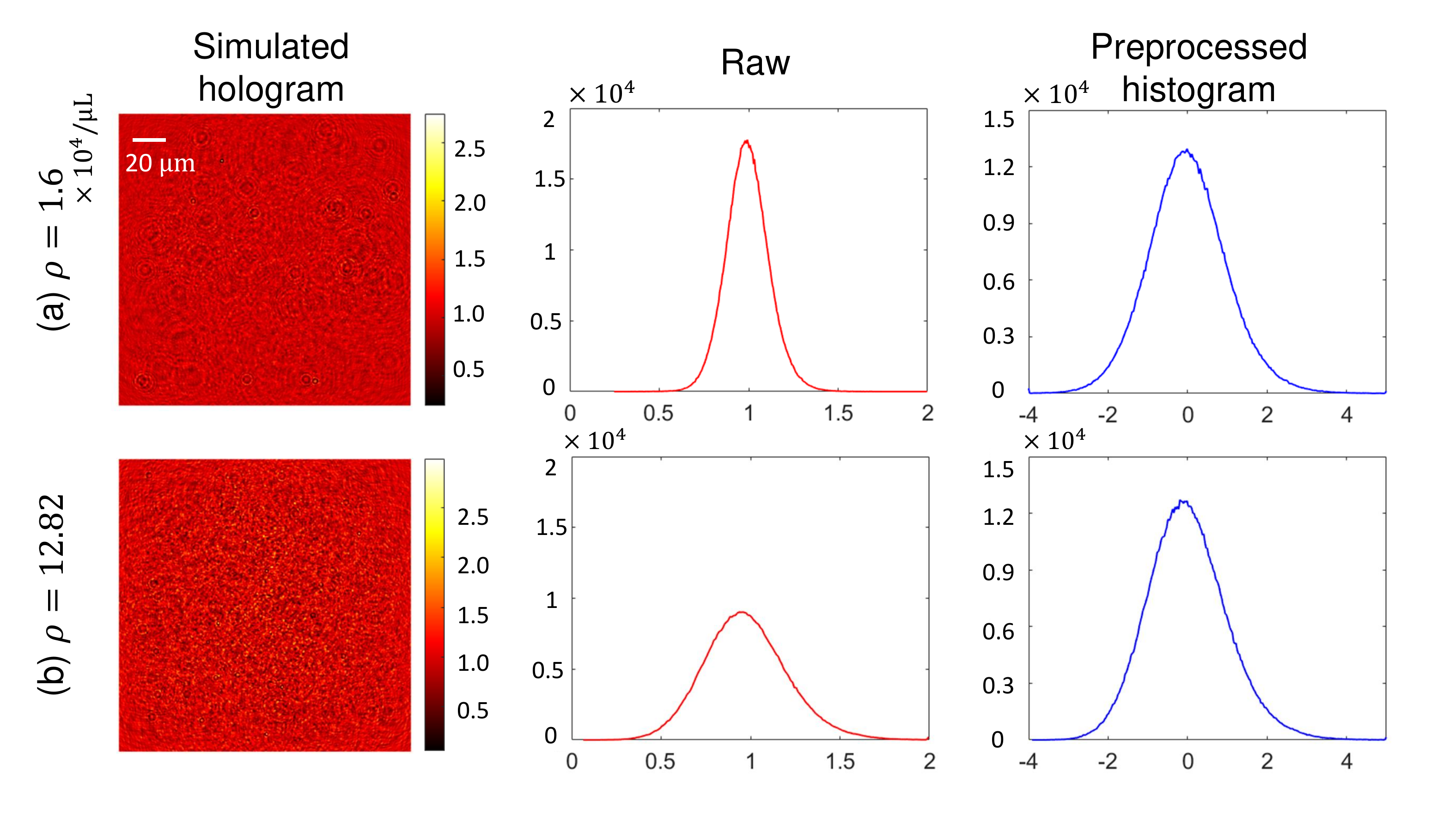}
	\caption{\textbf{The preprocessing result of holograms.} 
	(a) A simulated hologram at a low density ($\rho=1.6\times 10^4$~particles $\upmu \mr{L}^{-1}$), refractive index contrast 0.26, particle size $1~\upmu$m), the raw intensity histogram, and the histogram after preprocessing. 
	(b) A simulated holograms at a high density ($\rho=12.82\times 10^4$~particles $\upmu \mr{L}^{-1}$) particles, refractive index contrast 0.26, particle size $1~\upmu$m), the raw intensity histogram, and the histogram after preprocessing. 
	The raw histograms from the two particle densities are obviously different in both mean and standard deviation. 
	After preprocessing, both histograms approximately follow the same Gaussian distribution.}
	\label{fig:preprocessing}
\end{figure}
\begin{figure}
	\centering
	\includegraphics[width=1\linewidth]{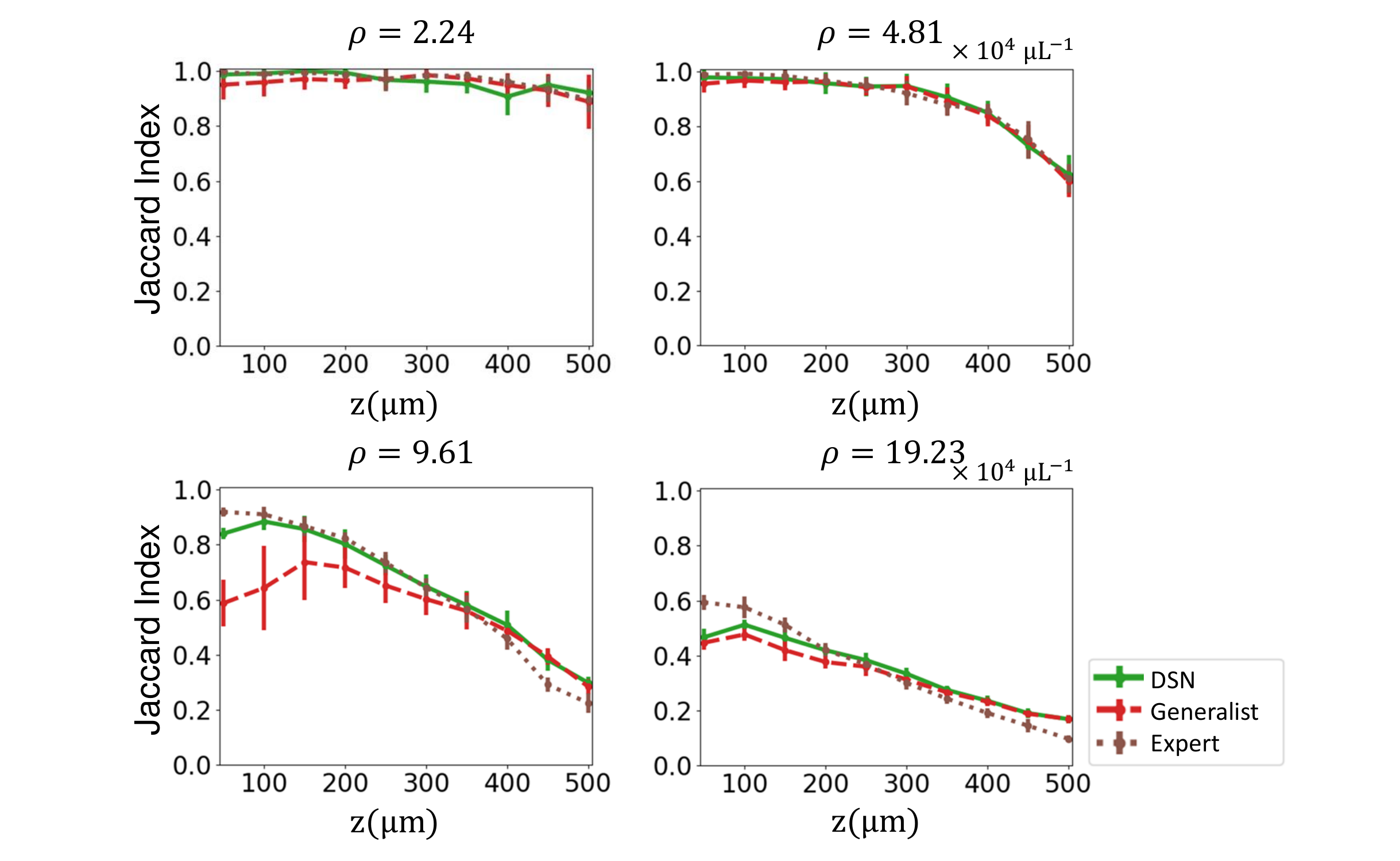}
	\caption{\textbf{Comparison between the DSN and the baseline generalist on unseen densities.} 
	Each subplot indicates the results on the testing data at the particle density labeled above each plot. 
	`Expert $\rho$' represents the expert DNN trained on the data with a particle density $\rho$ ($\times 10^{4}$ particles $\upmu \mr{L}^{-1}$).
	The DSN and the baseline generalist are trained using the same data from four other densities, as detailed in the main text.
	The DSN provides markedly higher accuracy than the generalist, in particular for high particle densities ($\rho \geq 9.61 \times 10^4$ particles $\upmu \mr{L}^{-1}$).
	}
	\label{fig:s_unseen_1X}
\end{figure}

\begin{figure}
	\centering
	\includegraphics[width=1\linewidth]{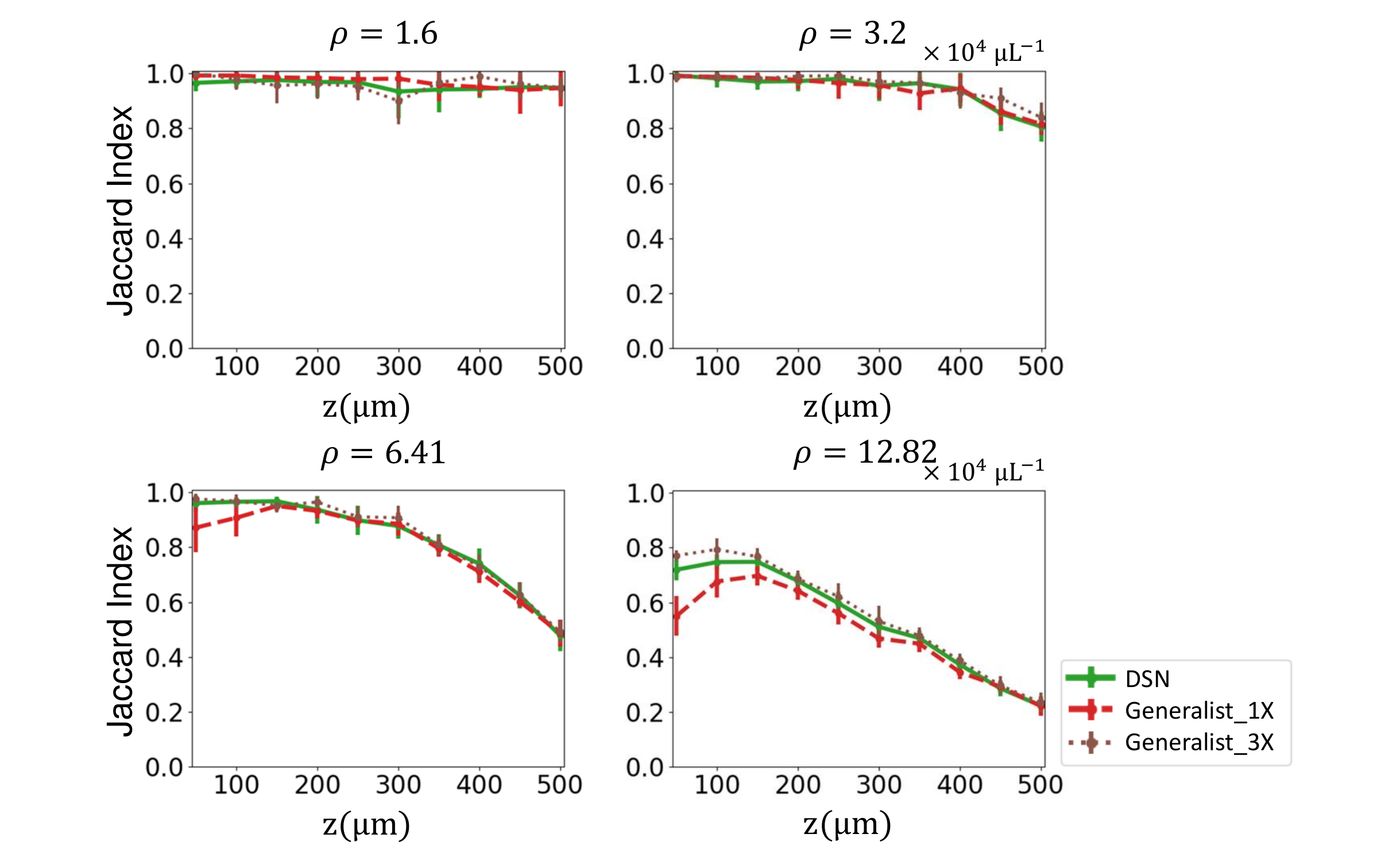}
	\caption{\textbf{Comparison between the DSN and the generalist networks on seen scattering densities.} 
	Each subplot indicates the results on the testing data at the particle density labeled above each plot with refractive index contrast 0.26 and particle size $1~\upmu$m. 
	The DSN, the baseline generalist (labeled as Generalist\_1$\times$) and the $3\times$ generalist (labeled as Generalist\_3$\times$) are trained using the same data from four other densities, as detailed in the main text.
	The DSN and the $3\times$ generalist  perform similarly in the  three lower densities.
	For the highest density, the $3\times$ generalist performs slightly better than the DSN at the shallow depths.
	}
	\label{fig:s_seen_3X}
\end{figure}

\begin{figure}
	\centering
	\includegraphics[width=0.92\linewidth]{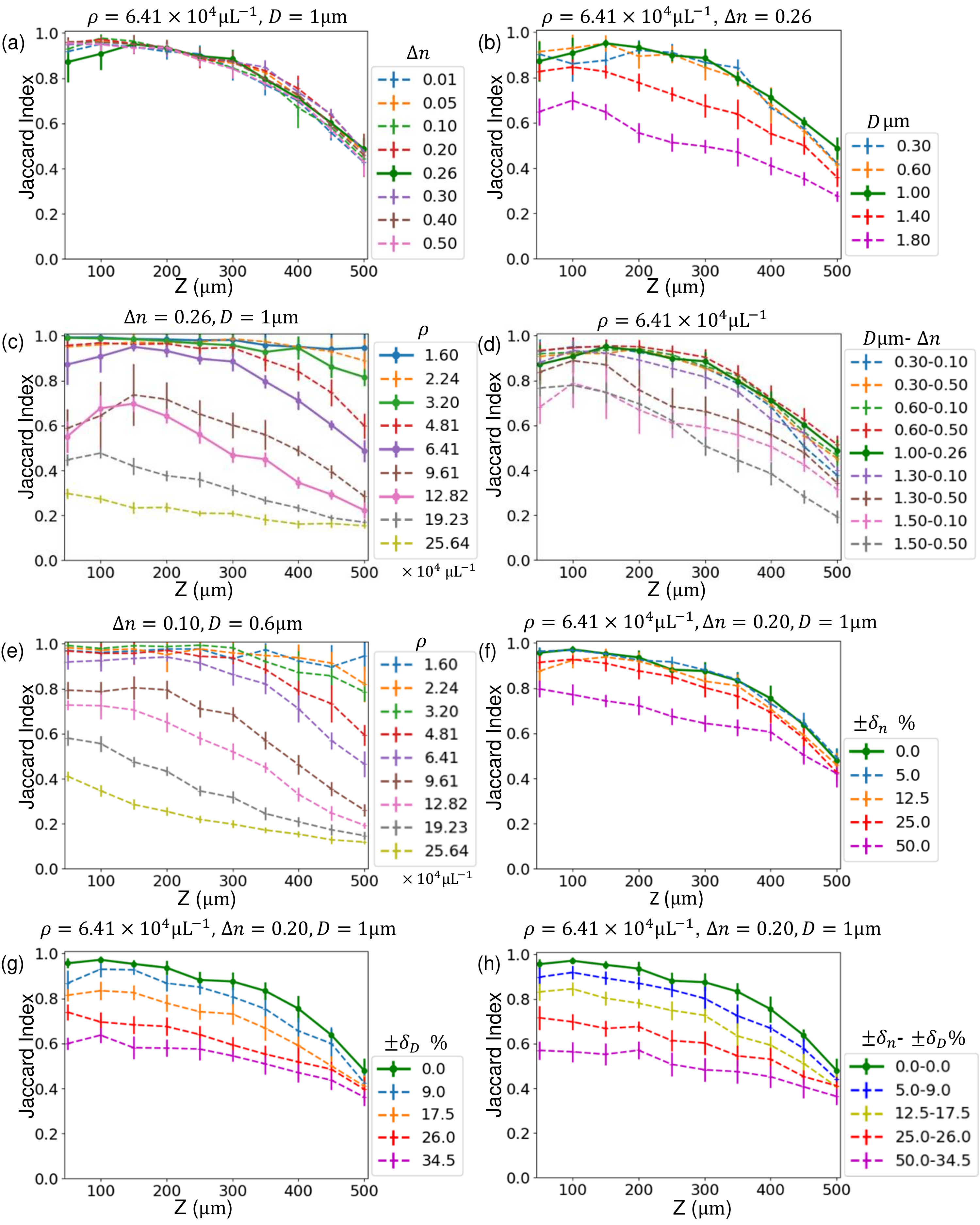}
	\caption{\textbf{Generalization of the baseline generalist to unseen scattering conditions.} 
The baseline seen cases are shown in solid lines; the ``unseen'' test conditions are in dashed lines. 
	The testing cases are identical to that for the DSN in Fig.~\ref{fig:unseen}, including:
	(a) unseen refractive index contrast;
	(b) unseen particle size; 
	(c) unseen particle density;
	(d) unseen refractive index contrast and particle size;
	(e) unseen refractive index contrast, particle size and density;
	(f) Uniformly distributed random refractive index contrast,  $\delta_n$\% denotes the variation range with respect to the central refractive index contrast;
	(g) Uniformly distributed random particle size, $\delta_d$\% denotes the variation range with respect to the central size; 
	(h) Uniformly distributed random refractive index and particle size. 
	In (f)-(h), the green dash-dotted line is the baseline unseen case at $\rho=6.41\times 10^{4}$~particles $\upmu \mr{L}^{-1}$, with a fixed refractive index contrast $\Delta n=0.20$ and a fixed particle size $D=1.0~\upmu$m. 
		}
	\label{fig:unseen_1Xgen}
\end{figure}

\begin{figure}
	\centering
	\includegraphics[width=0.92\linewidth]{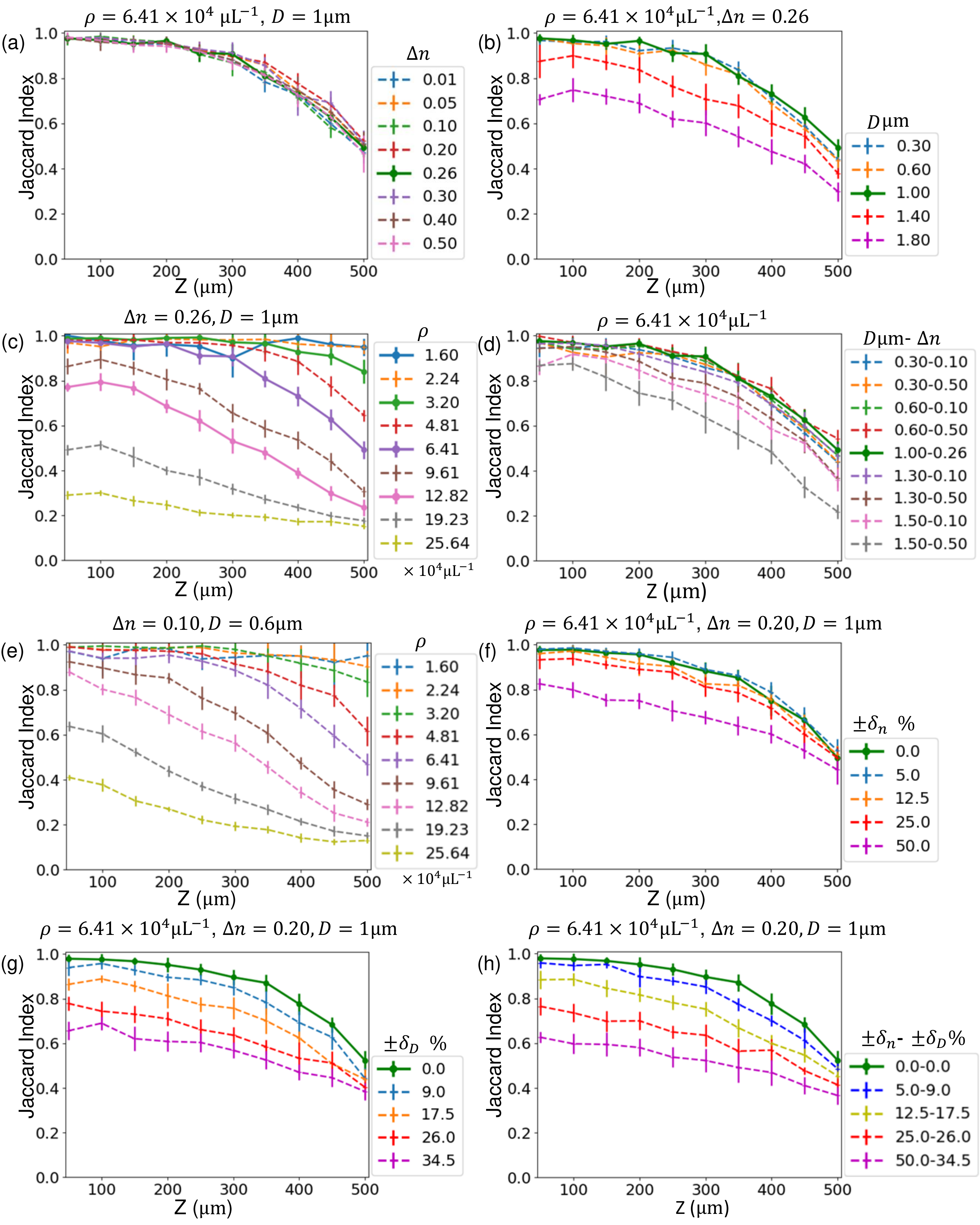}
	\caption{\textbf{Generalization of the $3\times$ generalist to unseen scattering conditions.} 
The baseline seen cases are shown in solid lines; the ``unseen'' test conditions are in dashed lines. 
	The testing cases are identical to that for the DSN in Fig.~\ref{fig:unseen}, including:
	(a) unseen refractive index contrast;
	(b) unseen particle size; 
	(c) unseen particle density;
	(d) unseen refractive index contrast and particle size;
	(e) unseen refractive index contrast, particle size and density;
	(f) Uniformly distributed random refractive index contrast,  $\delta_n$\% denotes the variation range with respect to the central refractive index contrast;
	(g) Uniformly distributed random particle size, $\delta_d$\% denotes the variation range with respect to the central size; 
	(h) Uniformly distributed random refractive index and particle size. 
	In (f)-(h), the green dash-dotted line is the baseline unseen case at $\rho=6.41\times 10^{4}$~particles $\upmu \mr{L}^{-1}$, with a fixed refractive index contrast $\Delta n=0.20$ and a fixed particle size $D=1.0~\upmu$m. 
		}
	\label{fig:unseen_3Xgen}
\end{figure}

\begin{figure}
	\centering
	\includegraphics[width=1\linewidth]{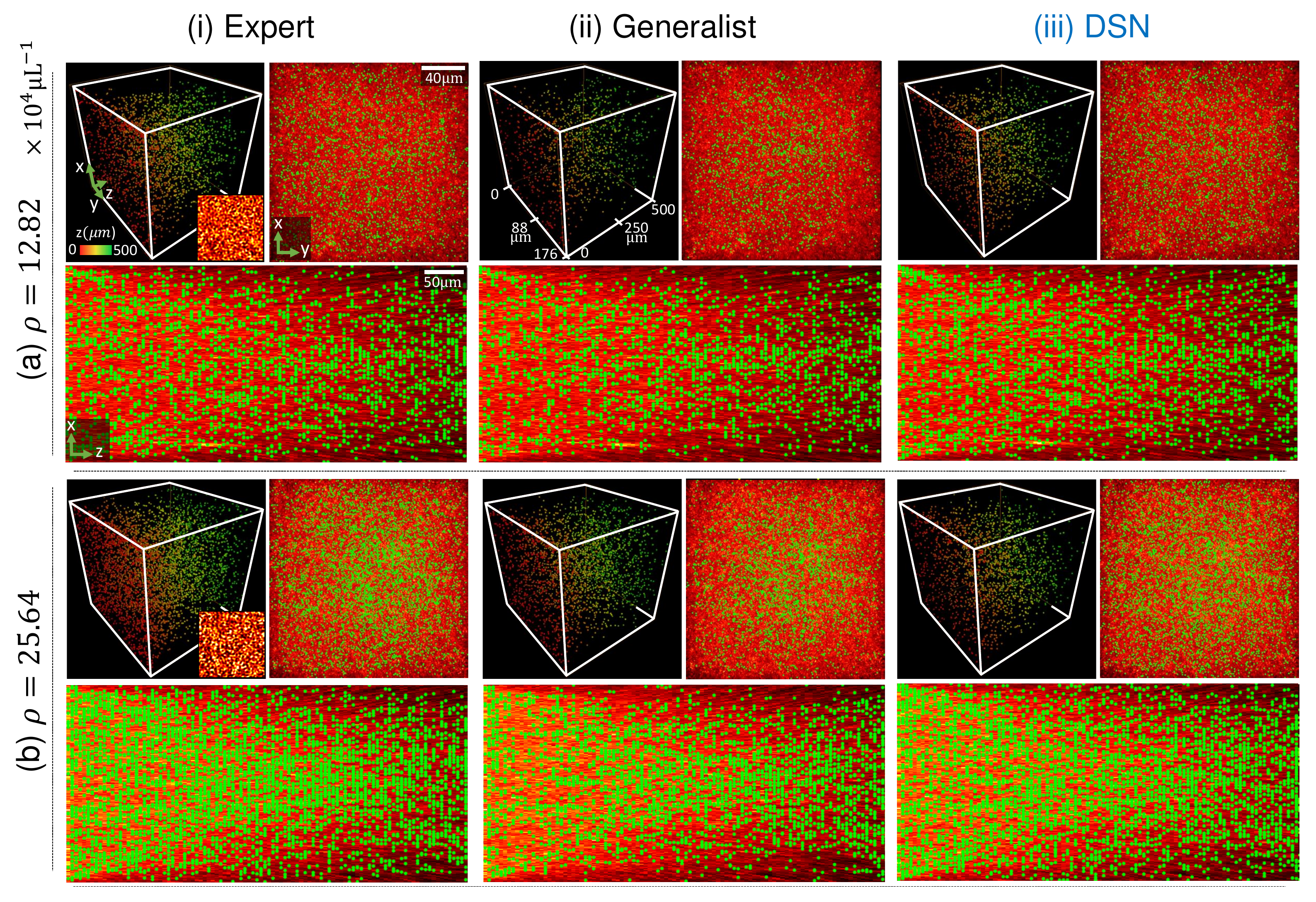}
	\caption{\textbf{Experimental results from the simulator-trained DNNs.} 
    	Particle 3D localization is shown for experimentally measured holograms using the simulator-trained (i) expert, (ii) generalist, and (iii) DSN networks at two higher particle densities. 
    	Each panel shows (Top left) the 3D rendering of the localization result with depth color-coded particles, with an inset showing a zoom-in of the measured hologram, (Top right) the maximum intensity $z-$ projection, and (Bottom) the $y-$projection of the DSN's 3D localization result (in green), overlaid on the respective $y-$ and $z-$projections of the corresponding holographic backpropagated volumes. 
		}
	\label{fig:s_exp}
\end{figure}
\clearpage

\renewcommand{\thetable}{S\arabic{table}}
\begin{table*}[p]
\centering
\caption{
	\textbf{The numbers of True Positive (TP), False Positive (FP), False Negative (FN)} (mean $\pm$ standard deviation) \textbf{for each particle number.}
	}
\begin{tabular}{ p{3cm} p{3cm} p{3cm} p{3cm}   }
 \hline
 \multicolumn{4}{c}{\textbf{Expert}} \\

 \textbf{Particle number}   & \textbf{TP}    &\textbf{FP} &   \textbf{FN}\\
 \hline
 \textbf{250} &   $247.2\pm2.9$  & $3.0\pm1.3$   &$2.8\pm2.9$\\

 \textbf{500} & $477.6\pm4.6$ & $3.9\pm3.2$  &  $22.4\pm4.6$ \\

 \rowcolor[HTML]{AAACED} \textbf{1000} & $863.4\pm14.4$ & $93.9\pm15.3$ & $136.6\pm14.4$\\

 \rowcolor[HTML]{AAACED} \textbf{2000} &  $1239.4\pm23.3$   & $518.0\pm46.3$  & $760.6\pm23.3$ \\
 \hline
 \multicolumn{4}{c}{\textbf{Generalist}} \\

  \textbf{Particle number}   & \textbf{TP}    &\textbf{FP} &   \textbf{FN}\\
 \hline
 \textbf{250} &   $248.5\pm2.3$  & $7.8\pm5.9$   &$1.5\pm2.3$\\

 \textbf{500} & $482.4\pm10.0$ & $21.3\pm8.2$  &  $17.6\pm10.0$ \\

 \rowcolor[HTML]{AAACED} \textbf{1000} & $864.3\pm16.8$ & $140.3\pm13.3$ & $135.7\pm16.8$\\

 \rowcolor[HTML]{AAACED} \textbf{2000} &  $1215.4\pm24.6$   & $630.4\pm27.3$  & $784.6\pm24.6$ \\
 \hline
 \multicolumn{4}{c}{\textbf{DSN}} \\

 \textbf{Particle number}   & \textbf{TP}    &\textbf{FP} &   \textbf{FN}\\
 \hline
 \textbf{250} &   $244.8\pm3.6$  & $13.7\pm6.0$   &$5.2\pm3.6$\\

 \textbf{500} & $479.4\pm6.2$ & $15.7\pm7.4$  &  $20.6\pm6.2$ \\

 \rowcolor[HTML]{AAACED} \textbf{1000} & $868.2\pm4.8$ & $65.7\pm11.5$ & $131.8\pm4.8$\\

 \rowcolor[HTML]{AAACED} \textbf{2000} &  $1246.3\pm23.2$   & $436.3\pm29.7$  & $753.7\pm23.2$ \\
 \hline
\end{tabular}
	\label{fig:s_count}
\end{table*}
\clearpage

\renewcommand{\thetable}{S\arabic{table}}
\begin{table*}[p]
\centering
\caption{\textbf{The kernel size and number of input, output channels of the Generalist network.}}
\begin{tabular}{ p{1.5cm} p{0.5cm} p{0.5cm} p{0.5cm} p{2.5cm} p{2.8cm} p{3cm}  }
 \hline
 \multicolumn{7}{c}{\textbf{Expert/Generalist}} \\

 \textbf{Layer number}   & \textbf{x}    &\textbf{y} &   \textbf{z} & \textbf{Input channel} & \textbf{Output channel} & \textbf{Total parameter}\\
\hline
 1 &  3  & 3  & 3 & 1& 16 & 448 \\

 2 &  3  & 3  & 3 & 16& 16 & 7168 \\

 3 &  2  & 2  & 2 & 16& 32 & 4608 \\

 4 &  3  & 3  & 3 & 32& 32 & 28672 \\

 5 &  2  & 2  & 2 & 32& 64 & 18432 \\

 6 &  3  & 3  & 3 & 64& 64 & 114688 \\

 7 &  2  & 2  & 2 & 64& 128 & 73728 \\

 8 &  3  & 3  & 3 & 128& 128 & 458752 \\

 9 &  2  & 2  & 2 & 128& 256 & 294912 \\

 10 &  3  & 3  & 3 & 256& 256 & 1835008 \\

 11 &  2  & 2  & 2 & 128& 256 & 294912 \\

 12 &  3  & 3  & 3 & 128& 128 & 458752 \\

 13 &  2  & 2  & 2 & 64& 128 & 73728 \\

 14 &  3  & 3  & 3 & 64& 64 & 114688 \\

 15 &  2  & 2  & 2 & 32& 64 & 18432 \\

 16 &  3  & 3  & 3 & 32& 32 & 28672 \\

 17 &  2  & 2  & 2 & 16& 32 & 4608 \\

 18 &  3  & 3  & 3 & 16& 16 & 7168 \\

 19 &  1  & 1  & 1 & 16& 1 & 32 \\
 \hline  
 \multicolumn{4}{c}{\textbf{Total}} &  \textbf{1233}  & \textbf{1713}  &  \textbf{3837408}\\
 \hline 

\end{tabular}
\label{fig:gen_parameters}
\end{table*}

\clearpage

\renewcommand{\thetable}{S\arabic{table}}
\begin{table*}[p]
\centering
\caption{\textbf{The kernel size and number of input, output channels of the $3\times$ Generalist.} 
}

\begin{tabular}{ p{1.5cm} p{0.5cm} p{0.5cm} p{0.5cm} p{2.5cm} p{2.8cm} p{3cm}  }
 \hline
 \multicolumn{7}{c}{\textbf{3X Generalist}} \\

 \textbf{Layer number}   & \textbf{x}    &\textbf{y} &   \textbf{z} & \textbf{Input channel} & \textbf{Output channel} & \textbf{Total parameter}\\
 \hline
 1 &  3  & 3  & 3 & 1& 28 & 784 \\

 2 &  3  & 3  & 3 & 28& 28 & 21952 \\

 3 &  2  & 2  & 2 & 28& 56 & 14112 \\

 4 &  3  & 3  & 3 & 56& 56 & 87808 \\

 5 &  2  & 2  & 2 & 56& 112 & 56448 \\

 6 &  3  & 3  & 3 & 112& 112 & 351232 \\

 7 &  2  & 2  & 2 & 112& 224 & 225792 \\

 8 &  3  & 3  & 3 & 224& 224 & 1404928 \\

 9 &  2  & 2  & 2 & 224& 448 & 903168 \\

 10 &  3  & 3  & 3 & 448& 448 & 5619712 \\

 11 &  2  & 2  & 2 & 224& 448 & 903168 \\

 12 &  3  & 3  & 3 & 224& 224 & 1404928 \\

 13 &  2  & 2  & 2 & 112& 224 & 225792 \\

 14 &  3  & 3  & 3 & 112& 112 & 351232 \\

 15 &  2  & 2  & 2 & 56& 112 & 56448 \\

 16 &  3  & 3  & 3 & 56& 56 & 87808 \\

 17 &  2  & 2  & 2 & 28& 56 & 14112 \\

 18 &  3  & 3  & 3 & 28& 28 & 21952 \\

 19 &  1  & 1  & 1 & 16& 1 & 32 \\
 \hline  
 \multicolumn{4}{c}{\textbf{Total}} &  \textbf{2145}  & \textbf{2997}  &  \textbf{11751408}\\
 \hline 
\end{tabular}
\label{fig:3_gen_parameters}
\end{table*}
\clearpage

\end{document}